\documentclass[prd,aps,nofootinbib,preprintnumbers, preprint, 12pt]{revtex4}
\usepackage{amsmath,amssymb,epsf}
\usepackage{graphicx, pstricks}

\def\be{\begin{equation}}
\def\ee{\end{equation}}
\def\ba{\begin{eqnarray}}
\def\ea{\end{eqnarray}}

\def\lta{~\mbox{\raisebox{-.6ex}{$\stackrel{<}{\sim}$}}~}
\def\gta{~\mbox{\raisebox{-.6ex}{$\stackrel{>}{\sim}$}}~}
\def\bq{\begin{quote}}
\def\eq{\end{quote}}

\def\brn#1{\left( #1\right)} 

\begin{document}


\title{ Gravity Waves Signatures  from Anisotropic  pre-Inflation}

\author{A.~E.~G\"umr\"uk\c{c}\"uo\u{g}lu$^{1}$, L.~Kofman$^{2}$ and  M.~Peloso$^{1}$}

\affiliation{$^1$ School of Physics and Astronomy, University of
Minnesota, Minneapolis, MN 55455, USA}
\affiliation{$^2$ CITA, University of Toronto, 60 St. George st.,
Toronto, ON M5S 3H8, Canada}

\date{\today}

\begin{abstract}

{We show that expanding or contracting Kasner universes are unstable due to the amplification of gravitational waves (GW).  As an application of this general relativity effect, we 
 consider a pre-inflationary anisotropic geometry characterized by a Kasner-like  expansion,
which is  driven dynamically towards inflation by a scalar field. We investigate  the evolution of linear metric fluctuations around this background, and calculate  the amplification of the long-wavelength  
GW of a certain polarization during the anisotropic expansion (this effect is absent for another GW polarization, and for scalar fluctuations).
  These GW are superimposed to the usual tensor modes of quantum origin from inflation, and are potentially observable if the total number of inflationary e-folds exceeds the minimum required to homogenize the observable universe only by a small margin.
Their contribution  to the temperature anisotropy  angular power spectrum decreases with the multipole $\ell$ as $\ell^{-p}$,  where  $p$ depends on the 
slope of the  initial GW power-spectrum. Constraints on the long-wavelength GW can be translated into limits on  the total duration of inflation
and the initial GW amplitude. The instability of  classical GW (and zero-vacuum fluctuations of gravitons) 
 during Kasner-like expansion (or contraction) may
 have other interesting applications. In particular, if GW become non-linear, they can significantly  alter the geometry before 
the onset of inflation.}

\end{abstract}

\preprint{arXiv:0807.1335}
\preprint{UMN-TH-2706/08}
\maketitle

\section{Introduction: Anisotropic pre-Inflation}

The inflationary stage of the very early universe  explains the dynamical
origin of  the observed isotropic and homogeneous FRW geometry.
The patch of the FRW geometry covers the cosmological horizon
and beyond if inflation lasted 
\begin{equation}\label{efold}
N=62-\ln \left(\frac{10^{16}GeV}{V^{1/4}} \right)+\Delta \ ,
\end{equation}
e-folds  or longer. Here $V$ is the potential energy of the inflation, and  $\Delta$ is a correction
from the (p)reheating stage after inflation, which is not essential for our discussion.
Chaotic inflationary models, associated with a large energy ($\sim$ GUT scale)  of $V^{1/4} \sim 10^{16}$GeV, predict a very large
 number of inflationary e-folds, $N \gg 62$. 
Long-lasting inflation erases all classical anisotropies and inhomogeneities of the pre-inflationary stage.
However, scalar and tensor vacuum fluctuations during inflation lead to almost scale free 
post-inflationary scalar and tensor metric inhomogeneities  around our   smooth   observable FRW patch. 

In particular, the amplitude of the gravitational waves generated from the vacuum fluctuations during inflation is proportional to
 $V^{1/2}$,  $h_k \simeq\frac{H}{2\pi M_p} \sim  \frac{V^{1/2}}{M_p^2}$ (where $M_p$ is the reduced Planck mass).  There are significant efforts to measure the $B$-mode of $\Delta T/T$ polarizations, since this  will provide a direct probe  of the scale of inflation.
The current $95 \%$ C.L. limits on $r$ (ratio of the tensor to scalar amplitudes of cosmological fluctuations) $r \lta 0.43$ (WMAP-only) and $ r \lta 0.2$ (WMAP plus acoustic baryon oscillation, plus supernovae) \cite{WMAP} shall be improved to $r \lta 0.1$ by the Planck mission \cite{planck}, to
$r \lta 0.05$ by the ${\rm C}_\ell$over \cite{clover}, EBEX \cite{ebex}, and Spider \cite{spider} experiments (see \cite{epic} for the study of a mission that can improve over these limits). While these limits imply a detection in the case of high energy inflation, a number of  other
inflationary models, including many of the string theory constructions have
lower energy, and therefore lead to GW of much smaller amplitude, which are virtually
unobservable through  $B$ mode polarization \footnote{Future gravitational waves astronomy may  allow to probe $r$ up to the level $10^{-6}$ with
BBO \cite{BBO} or ultimate DECIGO \cite{decigo} direct detection experiments.}.

In anticipation of the null signal observation of the primordial GW from inflation, it is worth  thinking 
about other implementations of this result for the theory of inflation, besides putting limits on the
energy scale $V^{1/4}$.  
There are models of inflation (including many string theory inflationary models) where 
the total number of e-folds,  $N$,  does not exceed the minimum   (\ref{efold}) by a large number. 
If the extra number of e-folds $\Delta N$ beyond   (\ref{efold})  is relatively small then 
pre-inflationary inhomogeneities of the geometry are not erased completely, and their residuals 
can be  subject to observational constraints.
In the context of this idea, in this paper we suggest an additional  mechanism to have observable gravitational waves associated with inflation.
 These gravitational waves are very different from the GW generated from the vacuum fluctuations during inflation. Firstly, they are the residual tensor inhomogeneities from the pre-inflationary stage. Secondly, they can be of a classical, rather than quantum, origin. Thirdly, while their initial amplitude  and spectrum are given by the initial conditions, they are significantly affected by the number of ``extra'' e-folds $\Delta N$. Therefore, observational limits on gravity waves result in constraints on a combination of $\Delta N$ and of the initial amplitude.

The choice of the initial geometry of the universe before inflation is wide open. In principle, one may assume an arbitrary geometry with significant tensor inhomogeneities  component, and much smaller scalar inhomogeneities. This choice is, however, very artificial. A much more comfortable choice of the pre-inflationary stage will be a generic anisotropic Kasner-like geometry with small inhomogeneities around it. The origin of the anisotropic universe with the scalar field can be treated with  quantum cosmology, or can be embedded in the modern context of the tunneling in the string theory landscape. In fact, a Kasner-like (Bianchi I) space
  was a rather typical choice in previous papers on pre-inflationary geometry, see e.g. \cite{pre}. Most of the works on an anisotropic pre-inflationary stage aimed to investigate how the initial anisotropy is diluted by the dynamics of the scalar field towards inflation \cite{starob83}. 

The formalism of linear fluctuations about an anisotropic geometry driven by a scalar field toward inflation was constructed only recently \cite{Gumrukcuoglu:2006xj,Pereira:2007yy,Gumrukcuoglu:2007bx,Pitrou:2008gk}. 
Besides the technical aspects of calculations of cosmological  fluctuations, there is a substantial conceptual  difference between computations in the standard inflationary setting and in the anisotropic case. For an isotropic space undergoing inflationary expansion, all the modes have an oscillator-like time-dependence at sufficiently early times, when their frequency coincides with their momentum. One can therefore  use   quantum initial conditions  for these modes.
  This is no longer the case for an expansion starting from an initial Kasner singularity. In this case, a range of modes, which can potentially be observed today
 (if $\Delta N$ is not too large), are not oscillating initially and therefore cannot be quantized on the initial time hyper-surface; as a consequence, there is an issue in providing the initial conditions for such modes.
For this reason  we will adopt another perspective, namely,
  we will consider generic small classical inhomogeneities around the homogeneous background, as an approximation to the 
  more  generic anisotropic and inhomogeneous cosmological solution.

Equipped with this philosophy, we consider an anisotropic expanding universe filled up by 
the  scalar field with a potential $V(\phi)$ which is typical for the string theory
inflation. We add generic linear metric fluctuations about this geometry. The evolution of these fluctuations is by itself an interesting academic subject.
 However, it acquires a special significance in the context of the GW signals from inflation, because of a new effect that we report here of amplification of 
long-wavelength GW modes during the Kasner expansion. 
This growth terminates when a mode enters the ``average'' Hubble radius (the average of that for all the three spatial directions), or, for larger wavelength modes, when the background geometry changes from anisotropic Kasner  to isotropic inflationary expansion. We perform explicit computations in the case of an isotropy of two spatial directions. In this case the computation becomes much more transparent and explicitly $k$ dependent. Fluctuations for arbitrary $a ,\, b ,\, c$ were considered in the formalism of
\cite{Pereira:2007yy,Pitrou:2008gk}, where the $k$ dependence is not explicit. We verified that our  results agree with
\cite{Pereira:2007yy,Pitrou:2008gk}
 in the axisymmetric $b=c$ limit. We find that only one  of the two GW polarizations undergoes significant  amplification. 
Therefore, even if we assume for simplicity equi-partition of the amplitudes of   the three inhomogeneous physical modes of the system (the scalar and the two GW polarization) at the initial time, the final spectra that will be frozen at large scales in the inflationary regime will be very different from each other, in strong contrast to what is obtained in the standard inflationary computations.

This result can have different consequences, that we explore in the present work. Suppose that the growing GW mode is still linear (but significantly exceeds other modes) when the space becomes isotropic. Then, we can have significant yet linear classical GW fluctuations at the beginning of inflation, say of amplitude $\lta {\cal O}(1)$. If the modes which correspond to the largest scales that we can presently observe left the horizon after the first $\Delta N$ e-folds of inflation, their amplitude decreased by the factor $e^{-\Delta N}$ in this period. If $\Delta N$ is relatively small, say $\sim {\cal O}(10)$ the freeze out amplitude of these GW modes would be $\sim {\cal O}(10^{-6})$. The angular spectrum of these GW will rapidly decrease  as  the multipole number $\ell$ grows, since smaller angular scales are affected by modes which spend more time inside the horizon during the inflationary stage. 

Suppose instead that the growing GW mode becomes non-linear before the onset of inflation. In this case the background geometry departs from the original onset. 

Besides the phenomenological signatures, it is interesting to study the origin of the amplification of the GW mode. It turns out that the effect of GW amplification is related to the anisotropic Kasner stage of expansion. Therefore we will separately study  GW in the pure expanding Kasner cosmology.
For completeness, we also include the study of GW in a contracting Kasner universe, which is especially interesting due to the universality of anisotropic Kasner approach to singularity. 

The plan of the paper is the following. In Section~\ref{sec:background} we discuss the evolution of the anisotropic universe driven by the scalar field towards inflation. In Section~\ref{sec-linear} we briefly review the formalism of the linear fluctuations in the case of a scalar field in an anisotropic geometry, paying particular attention to the GW modes. In Section~\ref{sec:decoupled} we compute the amplification of one of the two GW modes that takes place at large scales in the anisotropic era.
 In Section~\ref{sec:pair} we discuss instead the evolution of the other two physical modes of the system. In Section~\ref{sec:kasner} we study the evolution of the perturbations in a pure Kasner expanding or contracting Universe. In Section~\ref{sec:obs} we return to the cosmological set-up, and we compute the contribution of the GW polarization amplified during the anisotropic stage to the CMB temperature anisotropies. In particular, by requiring that the power in the quadrupole does not exceed the observed one, we set some limits on the initial amplitude of the perturbations  vs. the duration of the inflationary stage. In Section~\ref{sec:sum} we summarize the results and list some open questions following from the present study, which we plan to address in a future work.

\section{Background Geometry}\label{sec:background}

The anisotropic  Bianchi-I geometry is described by
\begin{equation}\label{bianchi1}
ds^2 = -dt^2 +a^2 dx^2 + b^2 dy^2+ c^2 dz^2 \ ,
\end{equation}
where $a(t), b(t), c(t)$ are the scale factors for each of the three spatial directions.

We consider a scalar field $\phi$ in this geometry. Many string theory inflationary models (for examples see \cite{K2LM2T,Conlon:2005jm}) have a very flat inflationary potential which changes abruptly around its minima. 
Therefore, to mimic this situation, we will use a simple inflaton  potential
\begin{equation}
V=V_0 \left(1- {\rm e}^{-\phi/\phi_0}\right)^2 \ ,
\label{poten}
\end{equation}
which has quadratic form $V \sim \phi^2$ around the minimum, and is almost flat $V \approx V_0$ away from it. To obtain the correct amplitude of scalar metric perturbations from inflation, we set
\begin{equation}\label{parameters}
 \frac{\phi_0}{M_p} = 10^{-3}\ ,  \quad\,\quad V_0^{1/4} = 10^{13} {\rm GeV} \ .
\end{equation}

The background dynamics is governed by the Einstein equations for the scale factors in the presence of
the effective cosmological constant $V_0$, plus a possible contribution from  the kinetic energy of the scalar field. Quantum cosmology or tunneling models of the initial expansion favor a small scalar field velocity.  Therefore we select the  small velocity initial conditions $\dot \phi^2 \ll V_0$. In this case the generic solutions of the Einstein equations with cosmological constant for $a, b, c$ are known analytically (see e.g. \cite{Pitrou:2008gk}) and can be cast as
\begin{equation}\label{factor}
\Big(a(t), b(t), c(t)\Big)=\left(a_{in}, b_{in}, c_{in}\right) \, \Big[\sinh (3H_0 t)  \Big]^{1/3} \, \left[\tanh \left(\frac{3}{2}H_0 t \right)
  \right]^{p_i-1/3}  \ ,
\end{equation}
where $p_i$ are the Kasner indices, $\sum_{i=1}^3p_i=1$, $\sum_{i=1}^3p^2_i=1$, and $H_0$ is the characteristic time-scale of isotropization by the cosmological 
constant, $H_0^2=\frac{V_0}{3M_p^2}$, while $a_{in}, b_{in}, c_{in}$ are the normalizations of the three scale factors

For earlier times $t \ll 1/H_0$ the anisotropic regime  is described by the vacuum  Kasner solution 
\begin{equation}\label{early}
\Big(a(t), b(t), c(t)\Big)=\left(a_0, b_0, c_0\right) \cdot t^{p_i} \ ,
\end{equation}
which corresponds to an overall expansion of the universe (the average scale factor $\big(abc \big)^{1/3}=t^{1/3}$ is increasing), although only two directions are expanding (two positive $p_i$-s) while the third one is contracting (the remaining $p_i$ is negative).

For later time  $t \gg 1/H_0$ the universe is isotropic and expanding exponentially
\begin{equation}\label{late}
\Big(a(t), b(t), c(t)\Big)=\left(a_0, b_0, c_0\right) \cdot e^{H_0 t} \ ,
\end{equation}
where the constant normalizations $a_0,\,b_0,\,c_0$ are typically chosen to be equal.

It is instructive to follow the evolution of the curvature in the model.
The Ricci tensor is (almost)  constant throughout the evolution 
up to the end of inflation
\begin{equation}\label{ricchi}
R^{\mu}_{\nu}=\frac{1}{4} \delta ^{\mu}_{\nu} \, R \ ,  \quad\,\quad R=12  H_0^2
\end{equation}
At earlier times the Weyl tensor -- i.e. the anisotropic component of the curvature tensor -- gives
\begin{equation}\label{weil1}
C^2 \equiv C^{\mu\nu\rho\sigma}C_{\mu\nu\rho\sigma} = - \frac{16 \, p_1 \, p_2 \, p_3}{t^4} \ ,
\end{equation}
and, for  $t \ll 1/H_0$, $\sqrt{C^2}$ 
is much bigger than the isotropic components (\ref{ricchi}). This is why initially 
the contribution from the effective cosmological constant is negligible, and 
 the vacuum Kasner solution (\ref{early}) is a good approximation.
In contrast, at later times  $t \gg 1/H_0$
\begin{equation}\label{weil2}
C^{\mu\nu\rho\sigma}C_{\mu\nu\rho\sigma} \sim e^{-6H_0 t} \ ,
\end{equation}
and the anisotropic part of the curvature becomes exponentially subdominant relative to
its isotropic part driven by the cosmological constant. This is an illustration of the isotropization of the cosmological expansion produced by the scalar field potential. The timescale for the isotropization is
$t_{\rm iso} \equiv 1 / H_0 \,$.

In the following Sections we will study the equations for the linear fluctuations around the background
(\ref{bianchi1}), (\ref{factor}). These equations become significantly simpler and more transparent
for the particular choice of an axi-symmetric geometry e.g. when $c=b$, and the metric is
\begin{equation}\label{special}
ds^2 = -dt^2 +a^2 dx^2 + b^2 \left(dy^2+dz^2\right)\ .
\end{equation}
While the effect we will discuss is generic, for simplicity we will adopt the simpler geometry (\ref{special}) rather than the general Bianchi-I space (\ref{bianchi1}). In this case, the early time solution is a Kasner background with indices
 $p_1=-1/3, p_2=p_3=2/3$ \footnote{There is another axisymmetric asymptotic Kasner solution, with indices $p_1 = 1 ,\, p_2 = p_3 = 0$. The solution is a very special one, since it is the only Bianchi-I model with cosmological constant that 
is regular at $t=0$   (as can be easily checked by computing the curvature invariants; e.g. $C^2 = 12 \, H_0^4 \,$ at $t=0 \,$). For $H_0 =0$, this space is actually Minkowski space-time in an accelerated frame. Due to its special nature, we disregard this solution in the present study.}.
Also, it will be useful to define an ``average'' Hubble parameter $H$ and difference $h$ between the expansion rates in $x$ and $y$ (or $z$) directions as
\begin{equation}\label{hubbles}
H \equiv \frac{H_a+2\,H_b}{3}\ , \quad\,\quad
h \equiv \frac{H_a-H_b}{\sqrt{3}} \ , \quad\,\quad H_a=\frac{\dot a}{a} \ ,\; H_b=\frac{\dot b}{b}
\end{equation}
At earlier times  $H_a\approx -\frac{1}{3t}, H_b\approx \frac{2}{3t}$, while, at late times, 
$H_a=H_b \approx H_0$. The equation for the homogeneous scalar field is
\begin{equation}\label{scalar}
\ddot \phi+ 3H \dot\phi + V_{,\phi}=0  \ .
\end{equation}
Since the value of $H$ is very large initially, $H \approx \frac{1}{3 t}$, the Hubble friction 
keeps the field (practically) frozen at  $\phi = \phi_{in}$ during the anisotropic stage. 
For $t > t_{\rm iso}$, the universe becomes isotropic, and it enters  a stage of
slow roll inflation until $\phi$ rolls to the minimum of its potential.

We also will use another form of the metric (\ref{special}), with the conformal time $\eta$.
There is ambiguity in the choice of $\eta$, related to possible different choices of the scale factors in its relation with the physical time. We will use the average scale factor
\begin{equation}\label{ave}
a_{\rm av} \equiv \left(ab^2\right)^{1/3} \ ,
\end{equation}
and define $\eta$   through
\begin{equation}
d t = \left( a \, b^2 \right)^{1/3} d \eta \ , 
\end{equation}
which, at early times, gives $\eta \propto t^{2/3}$.
In this variable, the line element (\ref{special}) reads
\begin{equation}\label{special2}
ds^2 =\left(ab^2\right)^{2/3}\left[ -d\eta^2 +\left(\frac{a}{b}\right)^{4/3} dx^2 +
 \left(\frac{b}{a}\right)^{2/3} \left(dy^2+dz^2\right)\right]\ .
\end{equation}
In the following, dot denotes derivative wrt. physical time $t$, and prime denotes derivative wrt conformal time. Moreover, we always denote by $H_a$ and $H_b$ the Hubble parameters with respect to physical time.

\section{Linear fluctuations}\label{sec-linear}

In the FRW universe with a scalar field there are three physical modes of linear fluctuations.
Two of them are related to the two polarizations $h_{+}$ and $h_{\times}$ of the gravitational waves,
and one to the scalar curvature fluctuations $v$ induced by the fluctuations of the scalar field $\delta \phi$.
All three modes in the isotropic case are  decoupled from each other. The
formalism for the linear fluctuations on a FRW background has been extended
to the Bianchi-I anisotropic geometry in 
\cite{Gumrukcuoglu:2006xj,Pereira:2007yy,Gumrukcuoglu:2007bx,Pitrou:2008gk}. Again, there are three physical modes;
however, in the general case of arbitrary $a,b,c$ the modes are mixed, i.e. their effective
frequencies in the bi-linear action $\delta^2 S$ are not diagonal, as it is the case in the isotropic limit.

In a special case of the axi-symmetric Bianchi I geometry (\ref{special})
one of the three linear  modes of fluctuations, namely, one of the gravity waves modes,
is decoupled from the other two. This makes the analysis of fluctuations much more transparent than the
general $a\neq b \neq c$ case. While the effects we will discuss, we believe, is  common for arbitrary $a,b,c$, we will consider linear fluctuations around the geometry (\ref{special}).  The computation follows the formalism of \cite{Gumrukcuoglu:2007bx}, where the reader is referred for more details.

The most general metric perturbations around  (\ref{special}) can be written as 
\begin{equation}
 g_{\mu \nu} =\left(\begin{array}{llll}
- a_{\rm av}^2 \left(1+2 \, \Phi\right) & a_{\rm av} \, a \, \partial_1 \chi
& a_{\rm av} \, b \, \left( B_{,i} + B_i \right)\\
& a^2 \left(1-2\,\Psi\right)  & a \, b \, \partial_1 \left( {\tilde B}_{,i} + {\tilde B}_i \right)\\
&& b^2 \left[ \left( 1 - 2 \, \Sigma \right) \delta_{ij} + 2 \, E_{,ij} + E_{(i,j)} \right]
\end{array} \right)\,.
\label{metric}                 
\end{equation}
where the indices $i,j=2,3$ span the $\left\{ y ,\, z \right\}$ coordinates of the isotropic $2$d subspace.
The above modes are divided into $2$d scalars ($\Phi,\,\chi,\,B,\,\Psi,\,{\tilde B} ,\, \Sigma ,\, E$) and $2$d vectors ($B_i ,\, {\tilde B}_i ,\, E_i$, subject to the transversality conditions $\partial_i B_i = \partial_i
{\tilde B}_i = \partial_i E_i  = 0$)~\footnote{Notice that transverse $2$d vectors have one degree of freedom; contrary to the $3$d case, there are no transverse and traceless $2$d tensors.} according to how these modes transform under rotations in the isotropic subspace. The two sets of modes are decoupled from each other at the linearized level. In addition, there is the perturbation of the inflaton field $\varphi = \phi + \delta \phi \,$, which is also a $2$d scalar.

The gauge choice
\begin{equation}
\delta g_{1i,\,2d{\rm s}} = \delta g_{ij} = 0 \,,
\label{ourgauge}
\end{equation}
corresponding to ${\tilde B} = \Sigma = E = E_i = 0 $, completely fixes the freedom of coordinate reparametrizations. It is convenient to work  with the Fourier decomposition of the linearized perturbations. We can therefore fix a comoving momentum ${\bf k}$, and study the evolutions of the modes having that momentum. Since modes with different momenta are not coupled at the linearized level, this computation is exhaustive as long as we can solve the problem for any arbitrary value of 
${\bf k}$. More precisely, we denote by $k_L$ the component of the momentum along the anisotropic $x$ direction, and by $k_T$ the component in the orthogonal $y-z$ plane. We denote by $p_L$ and by $p_T$ the corresponding components of the physical momentum. Finally, we denote by $k$ and $p$ the magnitudes of the comoving and physical momenta, respectively. Therefore, we have
\begin{equation}
k^2 = k_L^2 + k_T^2 \;\;\;,\;\;\; p^2 = p_L^2 + p_T^2 = \left( \frac{k_L}{a} \right)^2
+ \left( \frac{k_T}{b} \right)^2
\label{mom}
\end{equation}

To identify the physical modes, one has to compute the action of the system up to
the second order in these linear perturbations. One finds that the modes $\Phi, \, \chi ,\, B $, and  $B_i$,
corresponding to the $\delta g_{0\mu}$ metric perturbations, are nondynamical, and can be integrated out of the action. This amounts in expressing the nondynamical fields (through their equations of motion)
in terms of the dynamical ones, and in inserting these expressions back into the quadratic action. For instance, for the nondynamical $2$d vector mode one finds
\begin{equation}\label{b3}
B_i = \left( \frac{b}{a} \right)^{1/3} \, \frac{p_L^2}{p^2} \, \left( \frac{a}{b} \, {\tilde B}_i \right)'
\end{equation}
The analogous expressions for the $2$d nondynamical scalar modes can be found in 
\cite{Gumrukcuoglu:2007bx}.

In this way, one obtains an action in terms of the three remaining dynamical modes $\delta \phi ,\, \Psi ,\, {\tilde B}_i$. Once canonically normalized, these modes correspond to the three physical perturbations of the system. The canonical variables are
\begin{equation}
V \equiv a_{\rm av} \left[ \delta \phi + \frac{p_T^2 \, \dot{\phi}}{H_a \, p_T^2 + H_b \left( 2 p_L^2 + p_T^2 \right)} \Psi \right] \;\;,\;\;
H_+ \equiv \frac{\sqrt{2} \, a_{\rm av} \, M_p \, p_T^2 \, H_b}{H_a \, p_T^2 + H_b \left( 2 p_L^2 + p_T^2 \right)} \Psi 
\label{sca2d}
\end{equation}
and 
\begin{equation}\label{canon}
H_\times \equiv \frac{M_p}{\sqrt{2}} \, \frac{p_L}{p} \, a_{\rm av} \, 
\epsilon_{ij} \, p_i \, {\tilde B}_j
\end{equation}
where $\epsilon_{ij}$ is anti-symmetric and $\epsilon_{12} = 1 \,$ (we stress that $H_\times$ encodes only one degree of freedom, since, due to the transversality condition of the $2$d vector modes,  $p_i \, {\tilde B}_i  = 0$).

The dynamical equations for the modes $H_+$ and $V$ are coupled to each other, while that of the $H_\times$ mode is decoupled
\begin{eqnarray}
&& H_\times'' + \omega_\times^2 \, H_\times = 0 \nonumber\\
&& \left( \begin{array}{c} V \\ H_+ \end{array} \right)'' + 
\brn{ \begin{array}{cc}
\omega_{11}^2 & \omega_{12}^2 \\
\omega_{12}^2 & \omega_{22}^2
\end{array} }  \, 
\left( \begin{array}{c} V \\ H_+ \end{array} \right) = 0\, \ .
\label{evol}
\end{eqnarray}
The explicit expressions for the frequency matrix $\omega_{ij}, ( i,j=1,2)$ are rather tedious
and given in   \cite{Gumrukcuoglu:2007bx}. 

In the limit of isotropic background, $b \rightarrow a$, also the $2$d scalars decouple, and the frequencies become
\begin{eqnarray}
&& \omega_\times^2 ,\, \omega_{22}^2 \rightarrow k^2 - \frac{a''}{a} \nonumber\\
&& \omega_{11}^2 \rightarrow k^2 - \frac{z''}{z} \;\;\;,\;\;\; z \equiv \frac{a^2 \phi'}{a'} \nonumber\\
&& \omega_{12}^2 \rightarrow 0
\label{omegiso}
\end{eqnarray}
Therefore, the mode $V$ becomes the standard scalar mode variable \cite{musa}, associated to the curvature perturbation, while the modes $H_+$ and $H_\times$ are associated to the two polarizations of the gravitational waves. Also notice that without the scalar field there are two physical modes which, due to the residual $2$d isotropy, are decoupled.

\section{Decoupled tensor polarization}\label{sec:decoupled}

Since there are no vector sources, the $2$d vector system describes a polarization of a gravitational wave obeying the ``free field equations'' $\delta R_{\mu\nu}=0$ (which reproduce the equation (\ref{b3}) and the first of (\ref{evol})). As we now show, this mode undergoes an amplification, which does not occur for the other two modes. This can be understood from considering the frequency of this mode
\begin{equation}
\frac{\omega_{\times}^2}{a_{\rm av}^2} =p_L^2  +  p_T^2 + \frac{H_a^2-14 \,H_a H_b - 5 \,H_b^2}{9} + 
\frac{\dot{\phi}^2}{2 M_p^2} - \left( H_a - H_b \right)^2 \frac{p_T^2 \left( - 2 p_L^2 + p_T^2 \right)}{p^4}
   \ .
\label{vec2dminus}
\end{equation}

In the isotropic case, for which $\omega_\times^2$ is given by (\ref{omegiso}), each mode is deeply inside the horizon, $k / a \gg H$, at asymptotically early times. This is due to the fact that $H$ is nearly constant, while $k/a$ is exponentially large at early times. As a consequence, $\omega_\times \simeq k$ in the asymptotic past, and the mode oscillates with constant amplitude. In the present case instead
\begin{equation}
p_L \propto a^{-1} \propto \eta^{1/2} \approx t^{1/3} \;\;\;\;,\;\;\;\;
p_T \propto b^{-1} \propto \eta^{-1} \approx t^{-2/3} \;\;\;\;,\;\;\;\;
H_a ,\, H_b \propto \eta^{-3/2} \approx t^{-1}
\label{earlypH}
\end{equation}
at early times ($\eta  ,\, t  \rightarrow 0^+$). Namely, as we go backwards in time towards the initial singularity, the anisotropic direction becomes large, and the corresponding component of the momentum of the mode is redshifted to negligible values. On the contrary, the two isotropic directions
become small, and the corresponding component of the momentum is blueshifted. However, the magnitude of the two Hubble parameters increases even faster. Therefore, provided we can go sufficiently close to the singularity, the early behavior of each mode is controlled by the negative term proportional to the Hubble parameters in eq.~(\ref{vec2dminus}).

To be more precise, if we denote by $a_0$ and $b_0$ the values of the two scale factors at some reference time $\eta_0$ close to the singularity, we have
\begin{equation}
\omega_\times^2 = a_{\rm av}^2 \left( \eta \right) \left[ - \frac{2}{a_{\rm av}^2 \left( \eta_0 \right)} \, \frac{\eta_0}{\eta^3} + \frac{k_T^2}{b_0^2} \, \left( \frac{\eta_0}{\eta} \right)^2 + {\mathcal O } \left( \eta^0 \right) \right]
\label{omegaapproxearly}
\end{equation}
where the first term in the expansion comes from the terms proportional to $H_{a,b}^2$ 
in eq.~(\ref{vec2dminus}), while the second term from the component of the momentum in the isotropic plane (cf. the early time dependences with those given in eq.~(\ref{earlypH})). We see that the frequency 
squared is negative close to the singularity, so that the mode $H_\times$ experiences a growth \footnote{A tachyonic frequency does not necessarily imply a growth of the physical fluctuations. For instance, this does not happen  for the GW modes in the FRW geometry.}. In a pure Kasner geometry, the relations (\ref{earlypH}) hold at all times. Therefore, one would find $p_L \gg p_T \gg H_{a,b}$ at asymptotically late times. For brevity, we will loosely say that 
the mode ``enters the two horizons $H_{a,b}^{-1}$'' at late times; the meaning of this is simply that the frequency is controlled by the momentum in this regime, $\omega_\times^2 \simeq a_{\rm av}^2 \, p^2 > 0 \,$, and the mode $H_\times$ enters in an oscillatory regime.

This simple description is affected by two considerations: firstly, we do not set the initial conditions for the modes arbitrarily close to the singularity, but at some fixed initial time $\eta_0 \,$; Secondly, the geometry changes from (nearly) Kasner to (nearly) de Sitter due to the inflaton potential energy.
Consequently, there are three types of modes of cosmological size. I: Modes with large momenta start inside the two horizons at $\eta_0 \,$. They oscillate ($\omega_\times^2 > 0$) all throughout the anisotropic regime, and they exit the horizon during the inflationary stage. II: Modes with intermediate momenta, for which (\ref{omegaapproxearly}) is a good approximation at $\eta_0$. These modes enter the horizons, and start oscillating, at some time $\eta > \eta_0$ during the Kasner era; they exit the horizon later during inflation. III: Modes with small momenta, that are always outside the horizons, and never oscillate during the Kasner and inflationary regimes. 

These considerations are crucial for the quantization of these modes. We can perform the quantization only as long as $\omega_\times^2 > 0$, and the mode is in the oscillatory regime. As we mentioned, 
during inflation, this is always the case in the past. Moreover, the frequency is adiabatically evolving ($\omega' \ll \omega^2$), and one can set quantum initial conditions for the mode in the adiabatic vacuum. This procedure is at the base of the theory of cosmological perturbations, and results in a nearly scale invariant spectrum at late times, once the modes have exited the horizon and become classical. In the case at hand, we cannot perform this procedure for modes of small momenta / large wavelength (modes III above). If inflation lasts sufficiently long, such modes are inflated to scales beyond the ones we can presently observe, so that the inability of providing initial quantum conditions is irrelevant for phenomenology. However, if inflation had a minimal duration, this 
problem potentially concerns the modes at the largest observable scales.

Irrespectively of the value of the frequency, it is natural to expect that the modes possess some ``classical'' initial value at $\eta_0 \,$. In the following we discuss the evolution of the perturbations starting with these initial conditions. Although we do not have a predictive way to set these initial values, we can at least attempt to constrain them from observations. In addition, we should worry whether the growth of $H_\times$ can result in a departure from the Kasner regime beyond the perturbative level, in which case the background solution described in the previous Section may become invalid (we discuss this in Section \ref{sec:kasner}).

As long as the frequency is accurately approximated by (\ref{omegaapproxearly}) in the Kasner regime, 
the evolution eq. for the mode $H_\times$ is approximately solved by
\begin{equation}
H_\times = C_1 \: \sqrt{\eta} \, J_3 \left[2\,k_T \left(\frac{a_0}{b_0}\right)^{1/3} \sqrt{\eta_0\,\eta}\right]+ 
C_2 \: \sqrt{\eta} \, Y_3 \left[2\,k_T \left(\frac{a_0}{b_0}\right)^{1/3} \sqrt{\eta_0\,\eta}\right]
\label{solveccl}
\end{equation}
where $C_{1,2}$ are two integration constants, while $J$ and $Y$ are the Bessel and Neumann functions.  The first mode increases at early times (small argument in the Bessel function) $\propto \eta^2$, while the second one decreases as $\eta^{-1} \,$. We disregard the decreasing mode in the following computations, $C_2 = 0 \,$ (moreover, this mode diverges at the singularity).

\begin{figure}[h]
\centerline{
\includegraphics[width=0.6\textwidth,angle=-90]{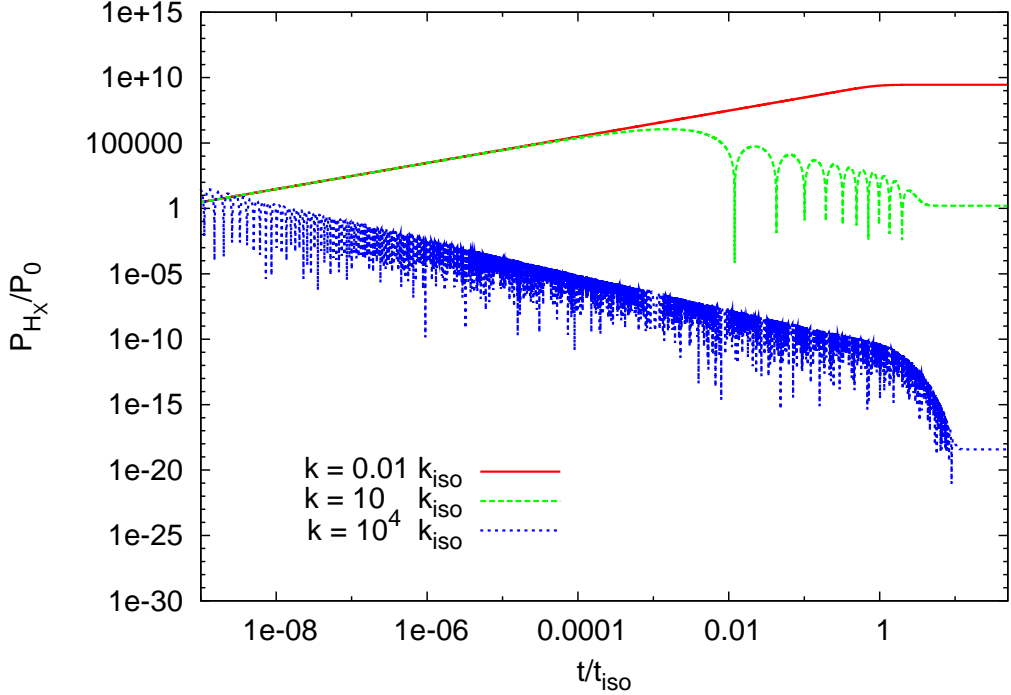}
}
\caption{Power in different modes of the decoupled tensor polarization, normalized to its initial value ($k_L = k_T $ for all the cases shown). We show the time evolution starting from 
$h_0 = - 10^6\sqrt{V_0} / M$ (defined in eq.~\ref{hubbles}). The universe isotropizes at $t_{\rm iso} = 1 / H_{\rm iso}$. The quantity $k_{\rm iso}$ is defined to be the momentum of the modes exiting the horizon at this time. Large scale modes do not exhibit oscillatory behavior, and grow during the anisotropic stage. Intermediate scale modes enter the horizons during the anisotropic era; this terminates their growth. Small scale modes are in the oscillatory regime all throughout the anisotropic phase, and do not experience any growth. All modes shown freeze during the inflationary stage. Notice the very different final values obtained in the three cases shown.}
\label{evoltcl} 
\end{figure}

\begin{figure}[h]
\centerline{
\includegraphics[width=0.4\textwidth,angle=-90]{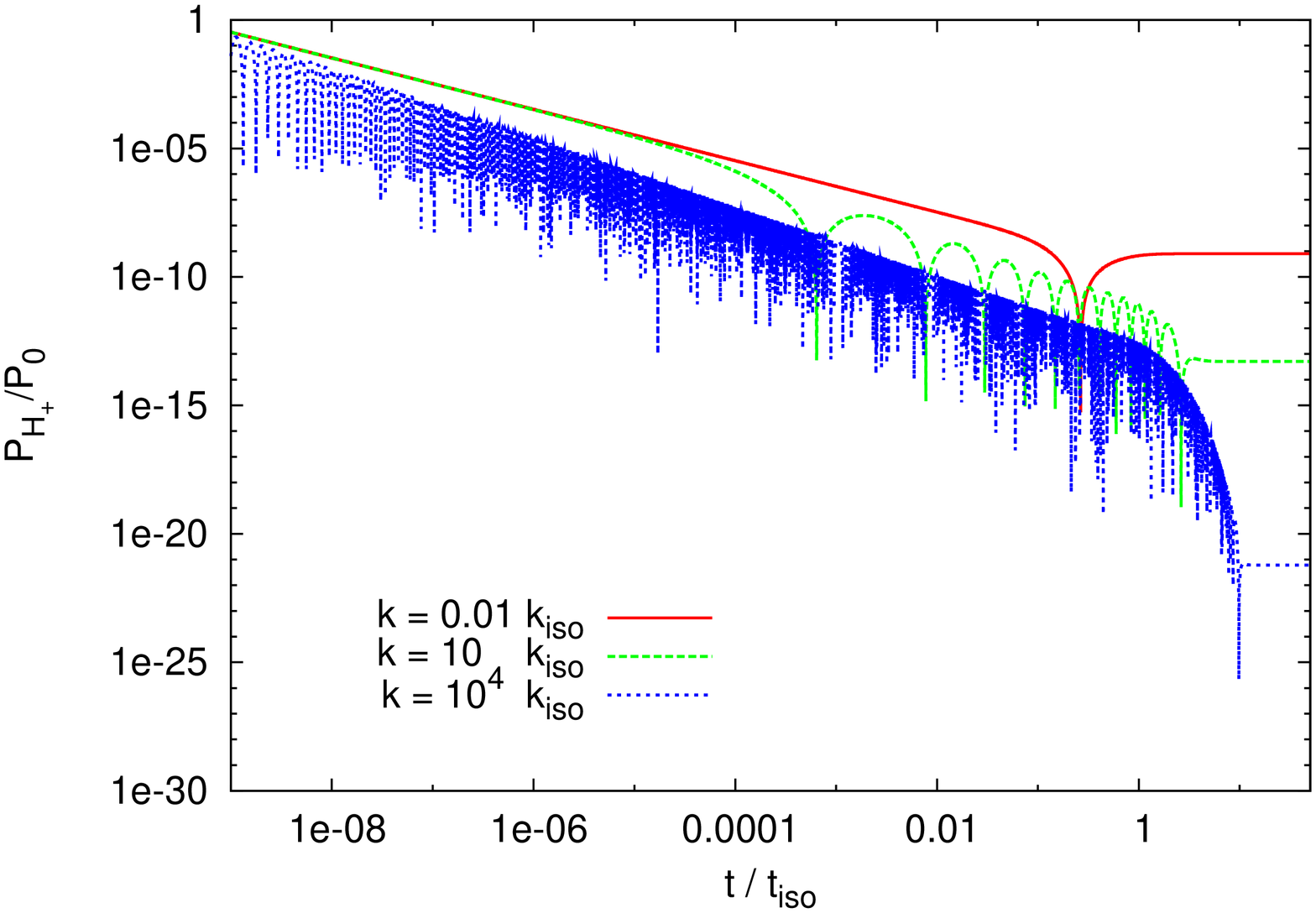}
\includegraphics[width=0.4\textwidth,angle=-90]{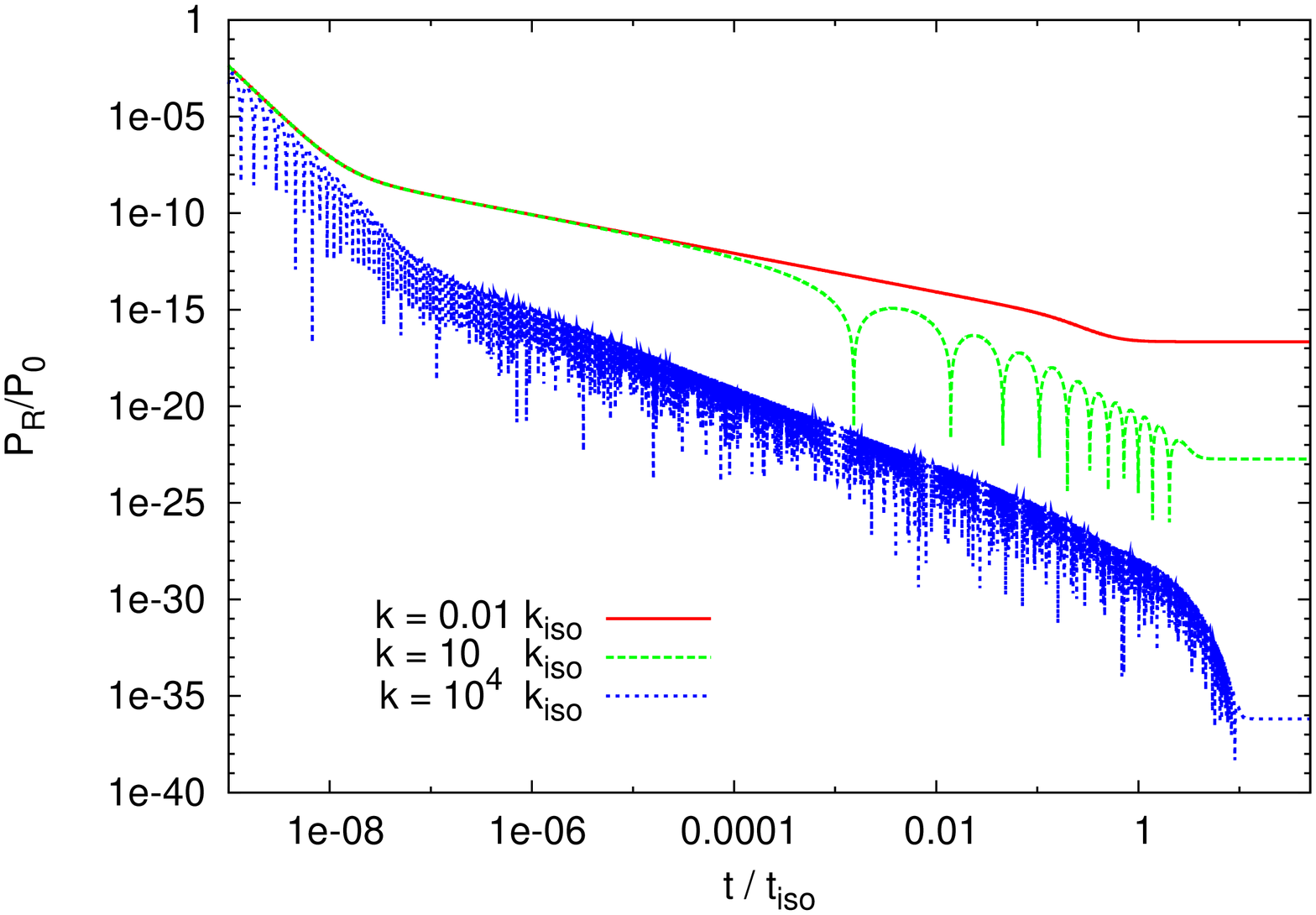}
}
\caption{Contribution of different modes to the power spectrum of the tensor polarization $H_+$  (left panel) and of the comoving curvature perturbation $R$  (right panel), normalized to its initial value. The background evolution, and the momenta shown, are the same as  in Figure \ref{evoltcl}. Contrary to the decoupled tensor mode, shown in Figure \ref{evoltcl}, the spectra do not grow while the modes are outside the horizons in the anisotropic regime.}
\label{evolpvcl} 
\end{figure}

Rather than the time evolution of $H_\times$, we show in Figure \ref{evoltcl} that of the corresponding power. We note the very different behavior obtained for large (III), intermediate (II), and small (I) scale modes. We choose to define the power  spectrum as
\begin{equation}
P_{H_\times} \equiv \frac{a_{\rm av}}{M_p^2 \, \pi^2} p^3 \vert H_\times \vert^2
\label{powtimes}
\end{equation}
This definition coincides with the standard one (see for instance \cite{Riotto:2002yw}) as the universe becomes isotropic. In particular, the power spectrum is frozen at large scales in the isotropic inflationary regime. Clearly, there is an ambiguity in this definition at early times, when the two scale factors differ (there is no a-priori reason for the choice of the average scale factor in this definition). This arbitrariness affects the time behavior shown in the Figure. However, it does not affect the relative behavior of the large vs. intermediate vs. small scale modes. Moreover, if we analogously define the power spectra for the $2$d scalar modes, the relative behavior of these two types of mode (Figure \ref{evoltcl} vs. Figure \ref{evolpvcl} ) is also unaffected by this arbitrary normalization. 

 The reality of this instability is demonstrated in Section \ref{sec:kasner}, where we compute the 
squared Weyl invariant due to these fluctuations (more precisely, we do so in an exact Kasner background, which, as we have remarked, coincides with the cosmological background at asymptotically early times).

\begin{figure}[h]
\centerline{
\includegraphics[width=0.6\textwidth,angle=-90]{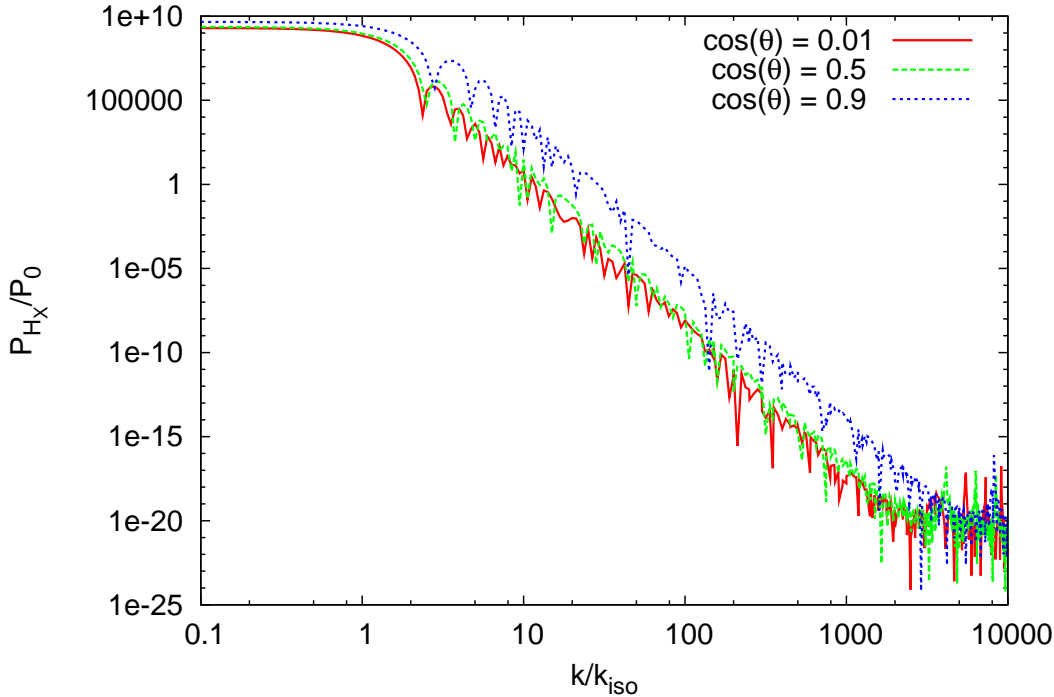}
}
\caption{Amplification of the power spectrum between the initial time (same as in the previous Figure) and some time during inflation for which the modes shown are outside the horizon, and the value of $P_{H_\times}$ is frozen. As explained in the main text, the growth is greater at small values of $k$ and $\theta$.}
\label{powertcl} 
\end{figure}

In Figure \ref{powertcl} we show the power spectrum (normalized to the initial value for each mode)
for the same background evolution as in the previous Figure, at some late time during inflation, when all the modes shown are frozen outside the horizon. As the approximate solution (\ref{solveccl}) indicates, the growth of $H_\times$ occurs as long as the transverse momentum $p_T$ is smaller than the Hubble rates $\vert H_{a,b} \vert \,$. Therefore, modes with smaller $k_T$ experience a larger growth. We denote by $\theta$ the angle between the comoving momentum, and the anisotropic direction,
\begin{equation}
k_L = k \, {\rm cos \theta} \;\;,\;\; k_T = k \, {\rm sin \theta}
\end{equation}
Therefore, in general, we expect a greater growth at smaller values of $k$ (for any fixed $\theta$) and at smaller values of $\theta$ (for any fixed $k$). \footnote{Notice that, due to the planar symmetry of the background, the same results are obtained at $\theta$ and $\pi - \theta \,$.} This behavior is manifest in Figure \ref{powertcl}. Modes with $k \ll k_{\rm iso}$ experience the same growth during the Kasner era (since the leading expression for $\omega_\times^2$ is independent of the momentum in this regime). 
Then, the modes shown in the Figure exhibit a very strong $\sim k^{-9}$ dependence for $k_{\rm iso} \lta k \lta {\rm few} \times 10^{3} \, k_{\rm iso} \,$. We also see an increase of the power as $\theta$ decreases. We stress that Figure \ref{powertcl} shows the contribution of each mode to the power spectrum normalized to the value that that contribution had at the initial time $\eta_0 \,$. Therefore, if the original spectrum has a scale, or an angular dependence, this will modify the final spectrum (for comparison, for the isotropic computation of modes with adiabatic quantum initial condition, $P_{H_\times} \propto k^3 \, \vert H_\times \vert^2 \propto k^2$ at early times).

\begin{figure}[h]
\centerline{
\includegraphics[width=0.6\textwidth,angle=-90]{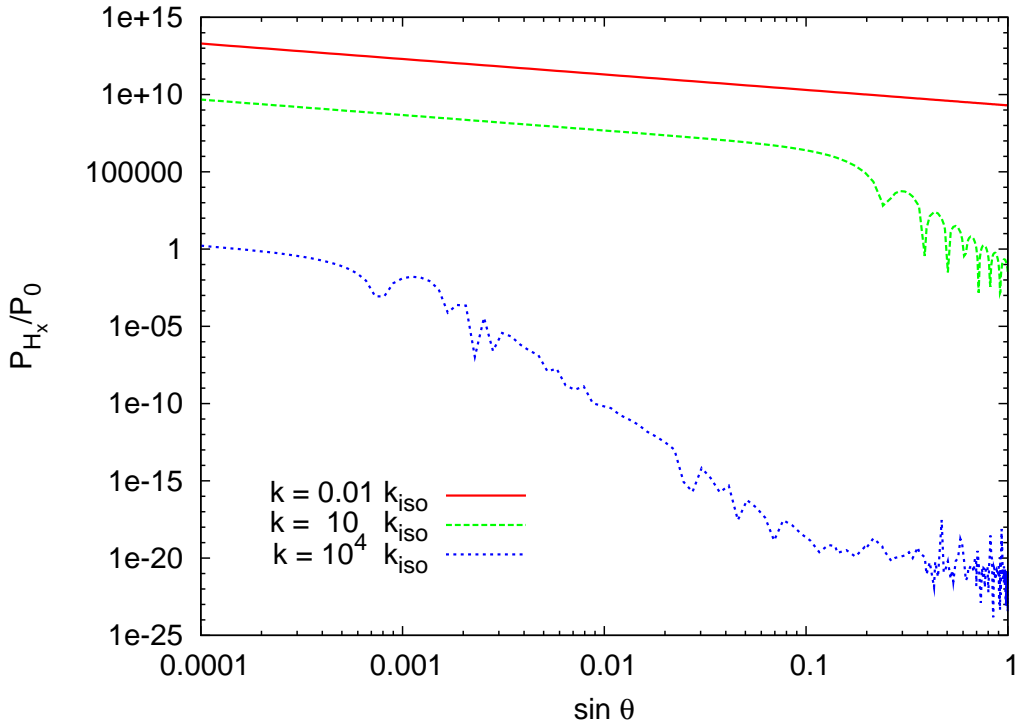}
}
\caption{Angular dependence of the power of different modes. We show the value reach by the mode once it is frozen in the inflationary stage, normalized to the initial value (the background evolution is identical to that of the previous two Figures). In the region $\theta \ll 1$ shown in the Figure ($k_T \ll k_L$) the final value exhibit a $1 / \sin \theta$ dependence.
}
\label{powerxi} 
\end{figure}

The large growth at small $\theta$ is more manifest in Figure \ref{powerxi}. The smallest angles shown in the Figure correspond to $k_T \ll k_L$, but to $p_T \gg p_L$ at the initial time (this is due to the different behavior of the two scale factors in the anisotropic era). In this region, the spectrum exhibits a milder $P_{H_\times} \propto k^{-3}$ dependence than for intermediate values of $\theta$. Finally, one may also consider smaller angles than those shown in the Figure, for which $p_T \leq p_L$ initially. We have found that final spectrum becomes $\theta-$independent in this region.

We conclude this section by discussing how the growth scales with the initial time. As long as $\vert H_{a,b} \vert \gg p_T \gg p_L$, the power  (\ref{powtimes}) of a mode grows as $\eta^{3/2}$ (as can be easily seen by combining the time dependences $a_{\rm av} \propto \eta^{1/2} ,\, p \simeq p_T \propto \eta ,\, H_\times \propto \eta^2 \,$). We use the initial value of $h$ (the difference between the two expansion rates, defined in eq. (\ref{hubbles})) as a measure on the initial time, since, contrary to the conformal time, this quantity is not affected by the normalization of the scale factors. Starting with a greater value of $\vert h_0 \vert$ corresponds to starting closer to the initial singularity, and, therefore, to a longer phase in which $P_{H_\times}$ grows. Since $h \propto t^{-1} \propto \eta^{-3/2} \,$ in the Kasner regime, the ratio $P_{H_\times} / P_0 \propto \vert h_0 \vert$ in the region in which the growth takes place. Although we do not show this here, we have verified that this scaling is very well reproduced by the numerical results.

\section{Coupled tensor polarization-scalar mode pair}\label{sec:pair}

As discussed in Section \ref{sec-linear} the two other physical modes of the system are coupled to each other in the anisotropic era. The evolution equations for the coupled system are formally given 
in (\ref{evol}). At early times, we find
\begin{eqnarray} 
\omega_{11}^2 , \omega_{22}^2 &=& a_{\rm av}^2 \left( \eta \right) \left[ \frac{1}{4 \, a_{\rm av}^2 \left( \eta_0 \right)} \, \frac{\eta_0}{\eta^3} + \frac{k_T^2}{b_0^2} \, \left( \frac{\eta_0}{\eta} \right)^2 + {\mathcal O } \left( \eta^0 \right) \right] \nonumber\\
\omega_{12}^2 = \omega_{21}^2 &=& a_{\rm av}^2 \left( \eta \right) {\mathcal O } \left( \eta^0 \right)
\label{omegaapproxearly+}
\end{eqnarray}
Therefore the coupling between the two modes can be neglected also at asymptotically early times. The main difference with the analogous expression for the decoupled tensor polarization, eq. (\ref{omegaapproxearly}), is that the squared eigenfrequencies of the two modes are now positive close to the singularity; therefore the two modes $V ,\, H_+$ do not experience the same growth as $H_\times \,$. Indeed, as long as the expressions (\ref{omegaapproxearly+}) are good approximations, we find the solution
\begin{equation}
H_+ = C_1^{H_+} \: \sqrt{\eta} \, J_0 \left[2\,k_T \left(\frac{a_0}{b_0}\right)^{1/3} \sqrt{\eta_0\,\eta}\right]+ 
C_2^{H_+} \: \sqrt{\eta} \, Y_0 \left[2\,k_T \left(\frac{a_0}{b_0}\right)^{1/3} \sqrt{\eta_0\,\eta}\right]
\label{solscacl}
\end{equation}
and an identical one for $V$ with the replacement $C_{1,2}^{H_+} \rightarrow C_{1,2}^{V} \,$ of the integration constants. Close to the singularity, the two modes grow as $\sqrt{\eta}$ and 
$\sqrt{\eta} \, {\rm ln } \, \eta \,$, respectively. Analogously to what we did for the $H_\times$ polarization, we disregard the mode which grows less at early times, $C_2^{H_+} = C_2^V = 0 \,$. 

We define the power spectra for the tensor polarization, and for the comoving curvature perturbation $R$ with the same prefactor as $P_{H_\times}$, cf. eq. (\ref{powtimes}),
\begin{equation}
P_{H_+} \equiv \frac{a_{\rm av}}{\pi^2} p^3 \vert H_+ \vert^2
\;\;\;,\;\;\;
P_R \equiv \left( \frac{H}{\dot{\phi}} \right)^2 \, \frac{a_{\rm av}}{2 \pi^2} \, p^3 \, \vert V \vert^2
\label{powepv}
\end{equation}
We see that, contrary to what happened for $H_\times \,$,  the coupled perturbations, and the corresponding power spectra, do not grow while outside the horizons during the anisotropic era.

\begin{figure}[h]
\centerline{
\includegraphics[width=0.4\textwidth,angle=-90]{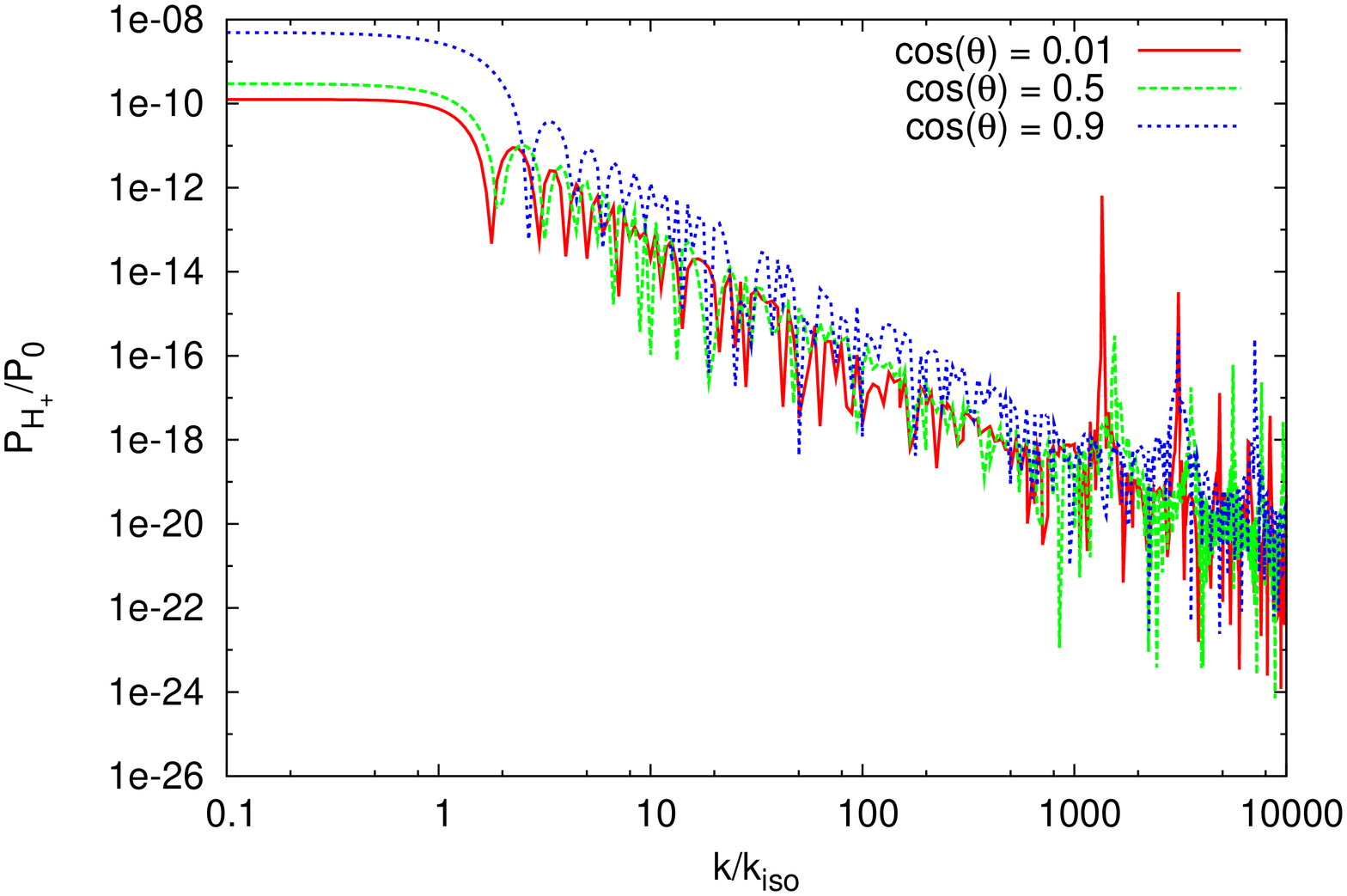}
\includegraphics[width=0.4\textwidth,angle=-90]{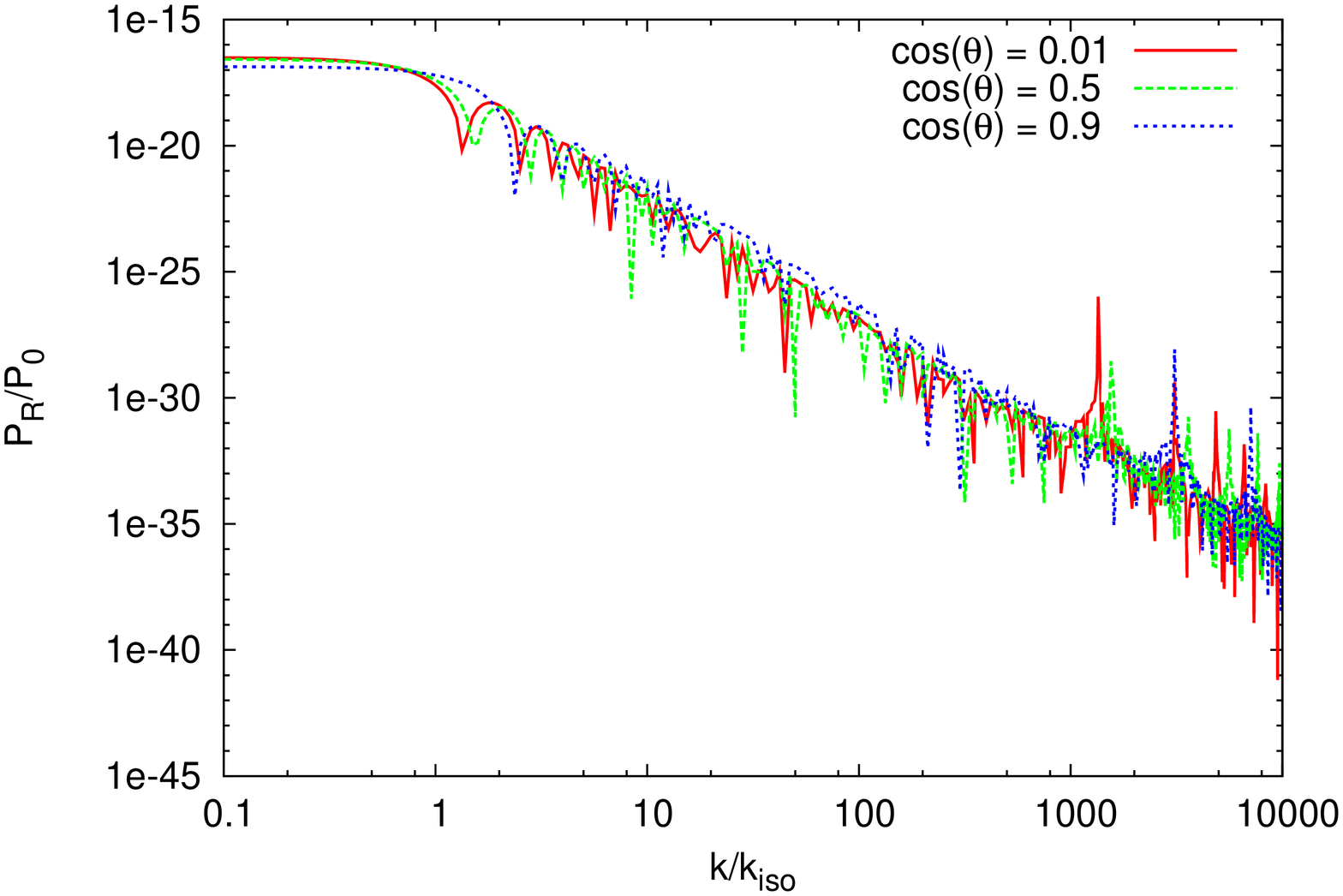}
}
\caption{Decrease of the power spectra of the tensor polarization $H_+$  (left panel) and of the comoving curvature $R$  (right panel). The Figure shows the ratio between the power spectra at some time during inflation (when the modes are outside the horizon in the isotropic regime, and the spectra are frozen), and the initial time. The decrease should be compared with the growth experienced at large scales by $P_{H_\times} \,$, shown in Figure \ref{powertcl}.}
\label{powerpscl} 
\end{figure}

This effect is also manifest in Figure \ref{powerpscl}, where we show the spectra of the tensor mode $H_+$ and the comoving curvature for the same range of momenta, and for the same angles $\theta$, as  those of $P_{H_\times}$ shown in Figure \ref{powertcl}.

\section{Instability of Kasner solution against Gravitational Waves} \label{sec:kasner}

The main result of the previous section was a significant amplification of  the mode $H_\times$, compared to the milder amplification of  the mode $H_+$, in the anisotropic background which is undergoing isotropization due to the effect of a scalar field. This growth can be ascribed to the instability of the Kasner geometry, either contracting or expanding, against gravitational waves which we are going to report  in this Section. Therefore in this Section we consider linearized gravity waves around an expanding and a contracting Kasner solution, without the presence of the scalar field, nor its fluctuation. In this case there are only the two decoupled modes $H_\times$ and $H_+ \,$.

The claim of  instability of the Kasner solution against the growth of the GW sounds at first glance heretic, at least for the contracting branch, in the light of the universality of the Belinskii-Khalatnikov-Lifshitz (BKL)  oscillatory regime of the Kasner epochs approaching the singularity. In fact, it is not, and, on the contrary, it is compatible with the BKL analysis. Moreover, our finding of the GW instability suggests a  new interpretation of the phenomena connected to the  instability, discussed by BKL and others for the contracting Kasner geometry in very different formalism and language \cite{BKL82,uggla,wain,Damour:2002et,Damour:2007nb}.

In this Section we first perform linearized calculations for  the classical gravitational waves around expanding and contracting Kasner solutions, and demonstrate their instabilities in terms of the evolution of the Weyl tensor invariant $C^{\mu\nu\rho\sigma}C_{\mu\nu\rho\sigma}$, which is independent of the gauge choice. We then connect the result with the  BKL analysis.

The background line element is given by equation (\ref{special2}), with the scale factors
\begin{equation}
a = a_0 \, \Big\vert \frac{\eta_0}{\eta} \Big\vert^{1/2}
\;\;\;,\;\;\; b = b_0 \, \Big\vert \frac{\eta}{\eta_0} \Big\vert
\label{abeta}
\end{equation}
This compact notation describes two disconnected geometries, at negative and positive conformal times, respectively. The algebraic expressions below simplify if we introduce the time $\eta_*$ in which the normalization of the two scale factors coincide. Therefore, rather than (\ref{abeta}), we can also use
\begin{equation}
a = a_* \, \Big\vert \frac{\eta_*}{\eta} \Big\vert^{1/2}
\;\;\;,\;\;\; b = a_* \, \Big\vert \frac{\eta}{\eta_*} \Big\vert
\label{abeta2}
\end{equation}
The two Hubble rates are
\begin{equation}
H_{a} = - \frac{1}{2\,a_*} \, \Big\vert \frac{\eta_*}{\eta} \Big\vert^{1/2} \, \frac{1}{\eta}
\;\;\;,\;\;\;
H_{b} = \frac{1}{a_*} \, \Big\vert \frac{\eta_*}{\eta} \Big\vert^{1/2} \, \frac{1}{\eta}
\label{hubconf}
\end{equation}
while the average scale factor is $a_{\rm av} = a_* \sqrt{\eta/\eta_*} \,$. The physical and conformal times are related by
\begin{equation}
d t = a_{\rm av} \, d \eta \;\;\;\Rightarrow\;\;\; t = \frac{2 \, a_*}{3} \, \Big\vert \frac{\eta}{\eta_*} \Big\vert^{1/2}
\, \eta
\end{equation}
For future use, we also define $t_*$ to be the physical time corresponding to $\eta_* \,$.

The solution with negative conformal time describes an overall contracting geometry, $a_{\rm av} \propto \left( - \eta \right)^{1/2} \,$, which crunches into the singularity at $\eta = 0$. The solution with positive conformal time describes instead an overall expanding space, $a_{\rm av} \propto \eta^{1/2}$, originating at the singularity at $\eta = 0 \,$. As in the previous Sections, we restrict the computation to the simpler case of a residual $2$d isotropy between two spatial directions. We expect that the instability occurs for general Kasner indices $\left( p_1 ,\, p_2 ,\, p_3 \right)$.

Now we turn to the linearized perturbations satisfying the vacuum equations $\delta R_{\mu\nu}=0$. 
There are two gravitational waves polarization perturbations. We consider a single mode with a given 
comoving momentum with components $k_L$ and $k_T$. We denote by $\vec{\xi}$ the vector in the $y-z$ plane along the direction of $\vec{k}_T$,
\begin{equation}
H \left( \eta ,\, {\vec x} \right) = {\rm e}^{-i k_L \, x - i \vec{k}_T \cdot \vec{\xi}}
H \left( \eta ,\, k_L ,\, k_T \right)
+ {\rm h. c.} \ .
\label{onemode}
\end{equation}

The two GW polarizations obey the equations
\begin{equation}
H_\times'' + \omega_\times^2 \, H_\times = 0 \;\;\;,\;\;\; H_+'' + \omega_+^2 \, H_+ = 0 \ ,
\label{eqomtomp}
\end{equation}
where the effective frequencies can be written in compact form
\begin{eqnarray}
\omega_{\times}^2 &=& 
a_{\rm av}^2 \left[ p^2 + H_a^2 \, \frac{p_L^4 + 20 p_L^2 \, p_T^2 - 8 p_T^4}{p^4} \right] \ , \nonumber\\
\omega_+^2 &=& 
a_{\rm av}^2 \left[ p^2 + H_a^2 \, \frac{16 p_L^4 + 296 p_L^2 \, p_T^2 + p_T^4}{\left( 4 p_L^2 + p_T^2 \right)^2} \right] \ ,
\label{omega}
\end{eqnarray}
both for the contracting and the expanding backgrounds. We recall that the physical momenta are related to the comoving one by the relations~(\ref{mom}).

After solving the two equations (\ref{eqomtomp}), we can compute the metric perturbations (\ref{metric})
and the Weyl tensor of the background plus perturbations. The square of the Weyl tensor, once expanded perturbatively,  has the following schematic  structure
\begin{equation}\label{pweyl}
C^{\mu\nu\rho\sigma}C_{\mu\nu\rho\sigma} = C^2 +C \delta C +\delta C^2  +\cdots \ ,
\end{equation}
where  $C^2$ is the Weyl invariant for the non-perturbed background solution, $C \, \delta C$ is the term linear in $H_{\times}$ and $H_+$, and $\delta C^2$ is the term quadratic in the perturbations. We do the computation for the two polarizations separately. For instance, for the $2$d vector modes, we compute the Weyl tensor in terms of the background and of the metric perturbations $B_3$ and ${\tilde B}_3$. We then relate the two perturbations to $H_\times$ through Eqs.~(\ref{b3}) and (\ref{canon}), expressing spatial derivatives in terms of the comoving momenta, see Eq.~(\ref{onemode}). In this way, we can write  the expression (\ref{pweyl}) in terms of $H_\times ,\, H_\times^*$, and their time derivatives. Finally, we insert in this expression the solutions of Eq.~(\ref{eqomtomp}). The procedure for the mode $H_+$ is analogous.

We compare the second and third term in (\ref{pweyl}) with the background term. A growth of the ratios
$C \, \delta C / C^2$ (denoted  as $\equiv \delta C / C$), or $\delta C^2/C^2$, signals an instability of the Kasner geometry. The background (zeroth order) Weyl invariant is
\begin{equation}
C^2 = \frac{12}{a_*^4} \, \frac{\eta_*^2}{\eta^6}=\frac{64}{27 \, t^4} \ .
\end{equation}
The first order term $C \, \delta C$ vanishes identically for the $H_\times$ polarization. It is nonzero for $H_+$, and it oscillates in space as  
${\rm exp} \left[ \pm \, i \left( k_L \, x + \vec{k}_T \cdot \vec{\xi}   \right) \right] \,$. The second order term $\delta C^2$ for the Fourier  modes $H_\times$ or  $H_+$ has a part which is constant in space, plus two parts that oscillate in space as 
 ${\rm exp}  \left[ \pm \, 2 i \left( k_L \, x + \vec{k}_T \cdot \vec{\xi}  \right) \right] \,$. In the following, we disregard the  oscillatory parts in $\delta C^2 \,$.

The solutions $H_\times$ and $H_+$ are either monotonically evolving in time, or oscillating, with an envelope that is monotonically evolving in time. The oscillatory regime takes place when the mode has a wavelength shorter than the Hubble radii, $p > \vert H_a \vert ,\, \vert H_b \vert$, and are absent in the opposite regime. Eqs.~(\ref{earlypH}) show the time dependence of the momentum and of the Hubble rates for the expanding Kasner solution. We see that, if we consider a complete background evolution, any mode starts in the large wavelength regime, and then goes in the short wavelength regime. Therefore, we expect that a mode is in the non-oscillatory regime sufficiently close to the singularity, and in the oscillatory regime sufficiently far from it (this behavior is manifest in the time evolutions shown). The same is true for the contracting background solution (with the obvious difference that 
early and late times interchange).

The time dependence of  the Weyl invariant (or of its amplitude, when the mode is oscillating) is summarized in Table~1. More specifically, we show the ratio between the terms proportional to the perturbations and the background one, both for the expanding and the contracting Kasner.
In the expanding case, a mode evolves from the large scale to the short scale regime. The opposite happens in the contracting case. Notice the the 
short scale behaviors in the expanding and contracting cases coincide. The same is not true for the two large scales behaviors. The reason is that, in the expanding case, we set to zero one of the two solutions of Eq.~(\ref{eqomtomp}) that is decreasing at early times, and that would diverge at the initial singularity.
The different behaviors are discussed in details in the next Subsection and in Appendix \ref{appA}.

\begin{table*}
\begin{tabular}{|c||c|c|c|c|}
\hline
 &  $\delta C / C $ for $H_+$    &
$\delta C^2 / C^2 $ for $H_+$
& $\delta C^2 / C^2 $ for $H_\times$ &
  \\
\hline \hline
$
\begin{array}{l}
{\rm EXPANDING }\\(0 < t < \infty)
\end{array}
$   & 
$\left\{ \begin{array}{c} {\rm const.} \\ {\rm const.} \end{array} \right.$
  & 
$\left\{ \begin{array}{c} {\rm const.}   \\ t^{16/3}   \end{array} \right.$
& $\left\{ \begin{array}{c} t^2 \\  t^4 \end{array} \right.$
& 
$  \begin{array}{l}
\leftarrow \; {\rm large \; scales} \\
\leftarrow \; {\rm short \; scales} 
\end{array}  $
\\ \hline
$
\begin{array}{l}
{\rm CONTRACTING }\\(- \infty < t < 0)
\end{array}
$   &
$\left\{ \begin{array}{c} {\rm const.}  \\ {\rm ln } \vert t \vert  \end{array} \right.$
  & 
$\left\{ \begin{array}{c} \vert t \vert^{16/3}  \\ \vert t \vert^{- 2/3}  \end{array} \right.$
& $\left\{ \begin{array}{c} \vert t \vert^4 \\ \vert t \vert^{-2} \end{array} \right.$
&
$  \begin{array}{l}
\leftarrow \; {\rm short \; scales} \\
\leftarrow \; {\rm large \; scales} 
\end{array}  $
\\
\hline
\end{tabular}%
\caption{
Summary table for the time dependence of $\delta C / C$ and of the (non-oscillating in space ) part of $\delta C^2 / C^2$. 
}
\label{tab1}
\end{table*}

\subsection{Contribution of the $H_\times$ mode to the Weyl invariant} \label{subHX}

In this Subsection, we compute the contribution of the $H_\times$ to the square of the Weyl tensor, both for an expanding and a contracting Kasner geometry. The analogous computation for the mode $H_+$ can instead be found in Appendix \ref{appA}.

\subsubsection{ $H_\times$ mode during expansion}

Plugging (\ref{abeta2}) into (\ref{omega}), the frequency has the large scales (early times) and short scales (late times) expansions
\begin{equation}
\omega_\times^2 \simeq
\left\{ \begin{array}{ll}
- \frac{2}{\eta^2} +  k_T^2 \, \frac{\eta_*}{\eta} & \,,{\rm ~large~scales}\\
 k_L^2 \, \left( \frac{\eta}{\eta_*} \right)^2 & \,,{\rm ~short~scales}\\
\end{array}\right.\,.
\label{esp-omX-exp}
\end{equation}

Consequently, we have the asymptotic solutions
\begin{equation}
H_\times \simeq
\left\{ \begin{array}{ll}
C_1 \, \sqrt{\eta} \, J_3 \left( 2 \,  \, k_T \, \sqrt{\eta_* \, \eta} \right)
& \,,{\rm ~large~scales}\\
\frac{{\bar C}_1}{\sqrt{\eta}} \, {\rm e}^{i \frac{  k_L \, \eta^2}{2  \, \eta_*}} + \frac{{\bar C}_2}
{\sqrt{\eta}} \, {\rm e}^{-i \frac{ k_L \, \eta^2}{2 \, \eta_*}}
& \,,{\rm ~short~scales}\\
\end{array}\right.\,.
\label{esp-hX-exp}
\end{equation}
where $C_1 ,\, {\bar C}_1 ,\, {\bar C}_2$ are three integration constants. In the early time solution we have disregarded a decaying mode that would diverge at $\eta \rightarrow 0 \,$, and where the expression in the second line is the large argument  asymptotic expansion of the parabolic cylinder functions
\begin{equation}
D_{-1/2} \left[ \sqrt{2} \, {\rm e}^{\pm i \, \pi /4} \, \ ,
 \sqrt{\frac{k_L}{\eta_*}} \, \eta \right]
\end{equation}
which are solutions of the evolution equation with the short scales expanded frequency (\ref{esp-omX-exp}). 

An extended computation of the square of the Weyl tensor (performed as outlined after Eq.~(\ref{pweyl})) leads to $C \, \delta C = 0 \,$;
 for the non-oscillatory part of the quadratic term in the fluctuations we find instead
\begin{equation}
\delta C^2  \simeq
\left\{ \begin{array}{ll}
\frac{2 \, \vert C_1 \vert^2 \, k_T^6 \, \eta_*^6}{3 \, M_p^2 \, a_*^6 \, \eta^3}
& \,,{\rm ~large~scales}\\
\left( {\bar C}_1 \, {\bar C}_2^* \, {\rm e}^{\frac{i  \, k_L \, 
\eta^2}{   \eta_*  }}
+ {\rm h.c.}  \right) \frac{32 k_L^4}{M_p^2
\, a_*^6 \, \eta_*}
& \,,{\rm ~short~scales}\\
\end{array}\right.\,.
\label{W-hX-exp}
\end{equation}

Since the background square Weyl $C^2 \propto \eta^{-6}$, this computation indicates that the ratio $\delta C^2 / C^2$ increases as $\eta^3$ in the large scales regime, while it oscillates in the short scales regime, with an amplitude that increases as $\eta^6 \,$. 

\begin{figure}[h]
\centerline{
\includegraphics[width=0.4\textwidth,angle=-90]{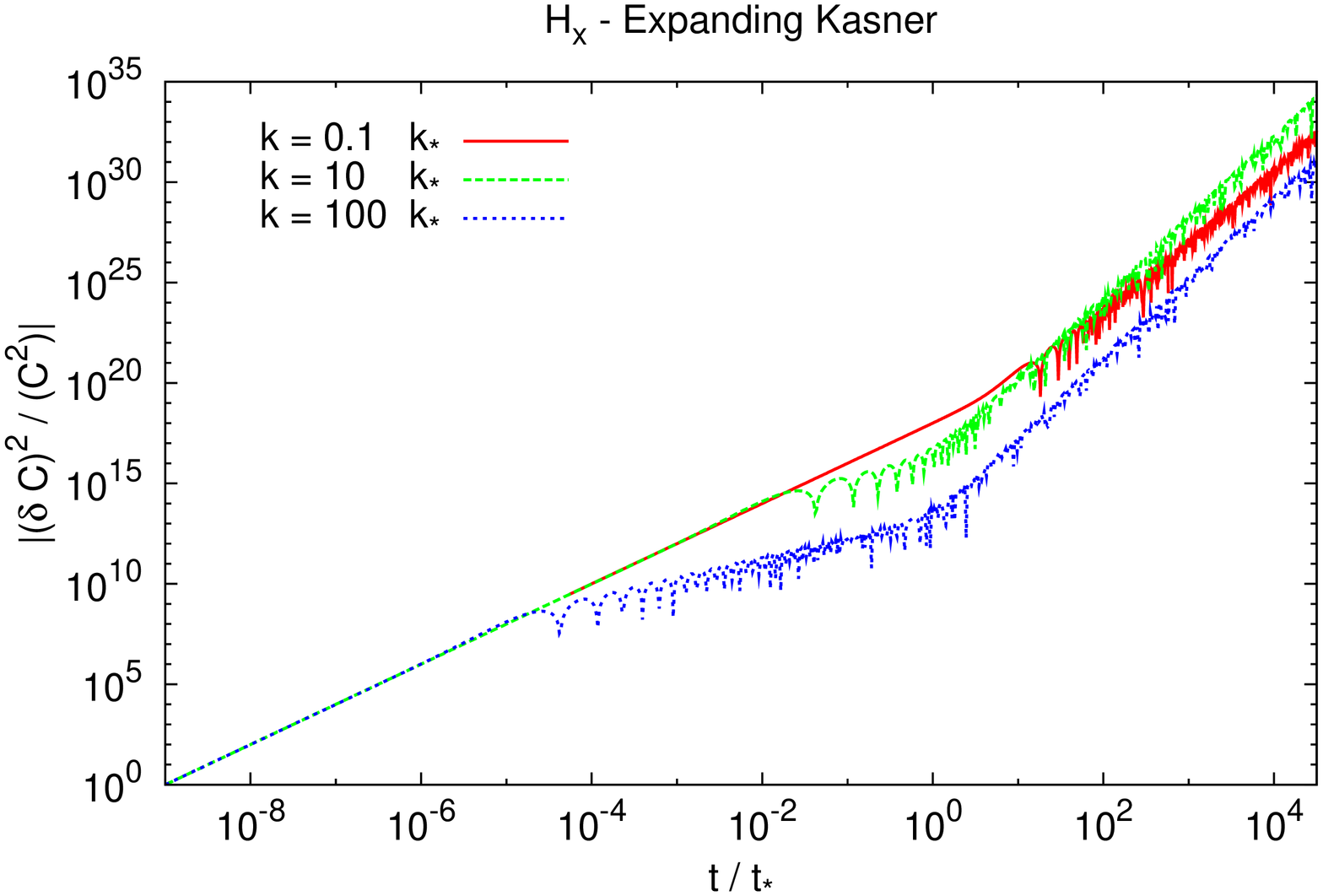}
\includegraphics[width=0.4\textwidth,angle=-90]{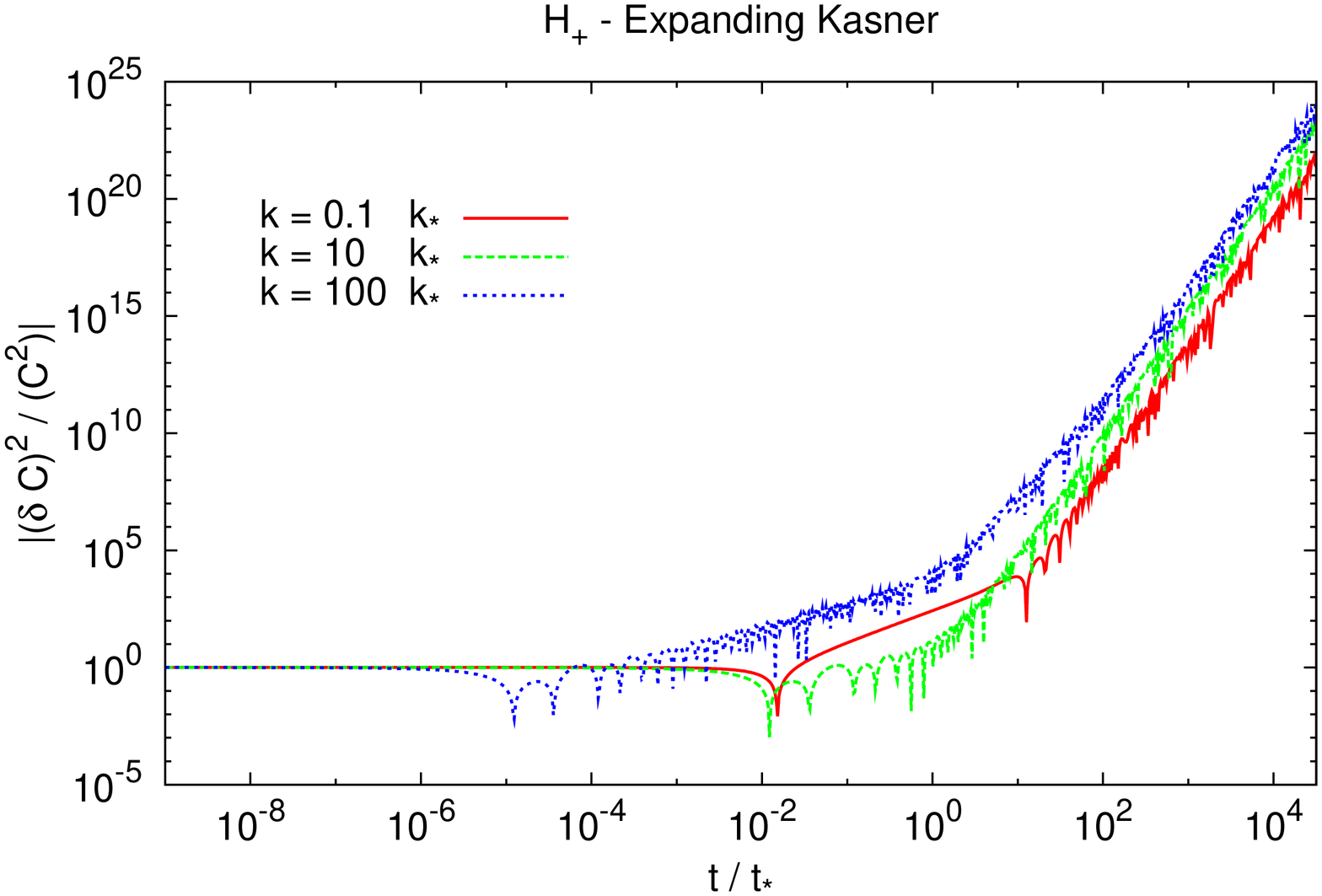}
}
\caption{
Contribution to $\delta C^2 / C^2$, defined in eq.~(\ref{pweyl}), from $H_\times$ (left panel) and 
$H_+$ (right panel) modes with some fixed comoving momentum. The time $t_*$ is a reference time at which the two scale factors have the same normalization. $k_*$ is the comoving momentum of the modes which have parametrically the same size as the two horizons at the time $t_*$ ($k_L = k_T$ in all the cases shown). See the discussions after eqs. (\ref{W-hX-exp}) and (\ref{W-hp-exp}) for more details.
}
\label{Weyl-exp} 
\end{figure}

In the left panel of Figure~\ref{Weyl-exp}  we present the full numerical evolution of $\delta C^2/C^2$ for three modes with different momenta. The behavior of the modes that we find here (pure Kasner geometry) should be compared with that discussed in the previous Sections, where the initial Kasner evolution was followed by an isotropic inflationary stage. In the evolutions shown in that case (for instance, Figs.~\ref{evoltcl} and \ref{evolpvcl} ) we had isotropization at the time $t_{\rm iso}$, and we started with an initial time of about $10^{-9} \, t_{\rm iso}$. In the present case, the geometry does not undergo isotropization. However, the two scale factors are normalized in such a way that they are equal to each other at the time $t_*$, cf. eqs. (\ref{abeta2}). Therefore, we also choose $t_0 = 10^{-9} \, t_*$ in the present case. Also in analogy to the modes shown in Figs.~\ref{evoltcl} and \ref{evolpvcl}, we define $k_*$ to be the comoving momentum of the modes which have parametrically the same size as the average horizon at the time $t_*$, namely $k_* / a_* = 1 / \left( 3 \, t_* \right) \,$ (cf. the expressions (\ref{hubconf})). Moreover, we choose $k_L = k_T$ as in those two Figures.

Figure~\ref{Weyl-exp}   confirms that each mode evolves from the non oscillatory large scales regime to the oscillatory short scales regime (the transition occurs at later times for modes of smaller momenta / larger scales). The time dependence of $\delta C^2/C^2$ shown in the Figure agrees with the one obtained analytically, and summarized in Table~\ref{tab1}.
 For comparison we also plot  in the Figure~\ref{Weyl-exp} the evolutions of the
$H_+$ mode during expansion considered in the Appendix A(a).

The results shown in the Figure confirm the instability of the background Kasner solution against the GW polarization $H_\times \,$. The growth in the large scales regime (early times) agrees with the amplification of the power spectrum shown in Figure~\ref{evoltcl}. However, contrary to what one would deduce from Figure~\ref{evoltcl}, we see that the growth continues also in the short scales (late times) regime. We recall that the definition of the power spectrum (\ref{powtimes}) contains an arbitrariness
in the overall time dependence (since one may have used a different combination of the two scale factors as overall normalization). We nonetheless adopted it to show the strong scale  dependence of the evolution of the power spectrum (which is not affected by the overall normalization), and the very different evolution experienced by the two GW polarizations (which is also independent of the arbitrary normalization, since $P_{H_\times}$ and $P_{H_+}$ are normalized in the same way). To properly study the instability, one must study invariant and unambiguous quantities, such as the (scalar) square of the Weyl tensor which is investigated in this Section.

\subsubsection{ $H_\times$ mode during contraction}

We consider now the contribution to the square of the Weyl tensor from the mode $H_\times$ in a contracting Kasner geometry. Plugging (\ref{abeta2}) into (\ref{omega}), the frequency of the $H_\times$ mode on the contracting background has the short and large scales expansions
\begin{equation}
\omega_\times^2 \simeq
\left\{ \begin{array}{ll}
k_L^2 \, \left( \frac{- \eta}{- \eta_*} \right)^2 & \,,{\rm ~short~scales}\\
- \frac{2}{ \left( - \eta \right)^2} + k_T^2 \, \frac{- \eta_*}{- \eta} 
& \,,{\rm ~large~scales}\\
\end{array}\right.\,.
\label{esp-omX-com}
\end{equation}

\begin{figure}[h]
\centerline{
\includegraphics[width=0.4\textwidth,angle=-90]{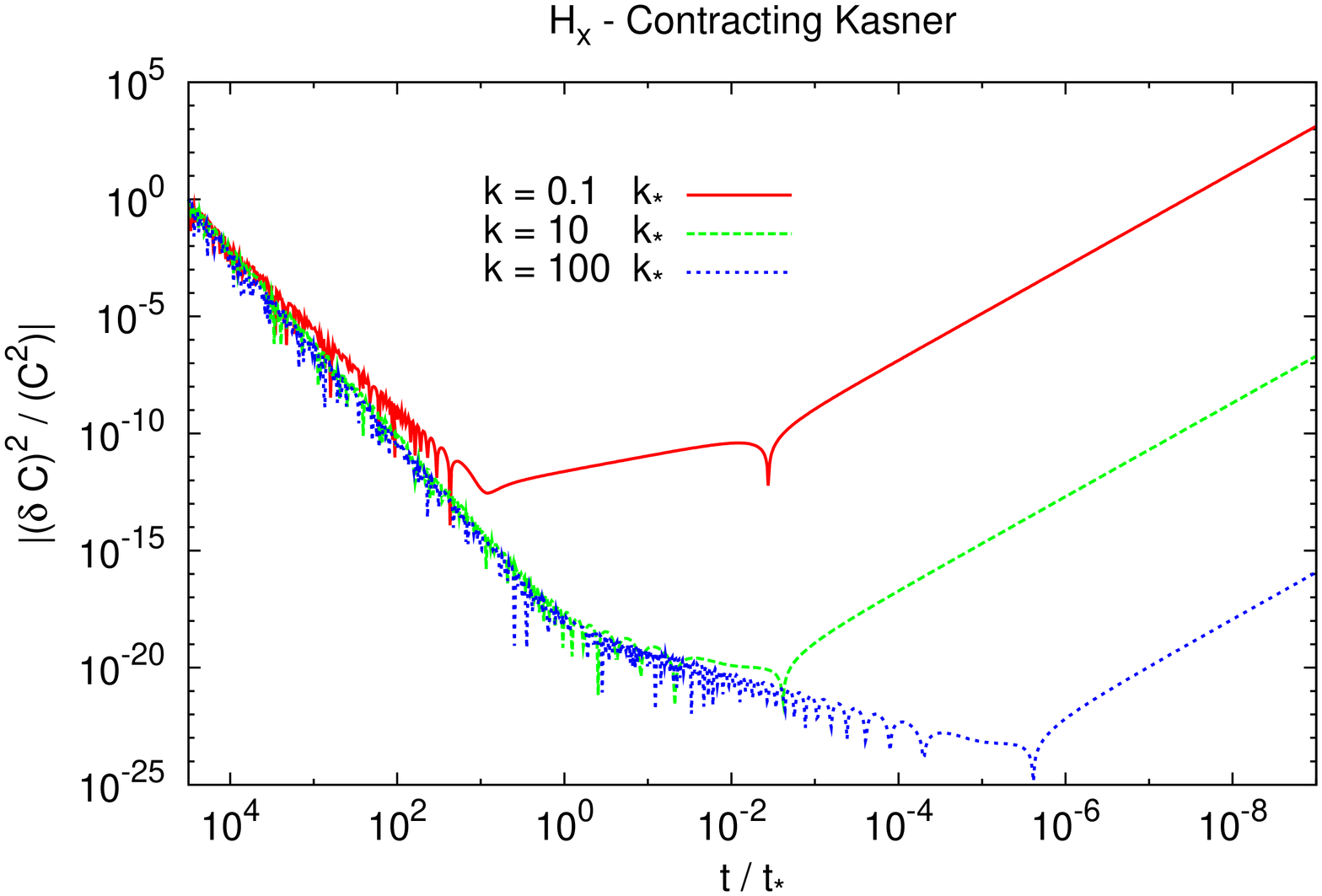}
\includegraphics[width=0.4\textwidth,angle=-90]{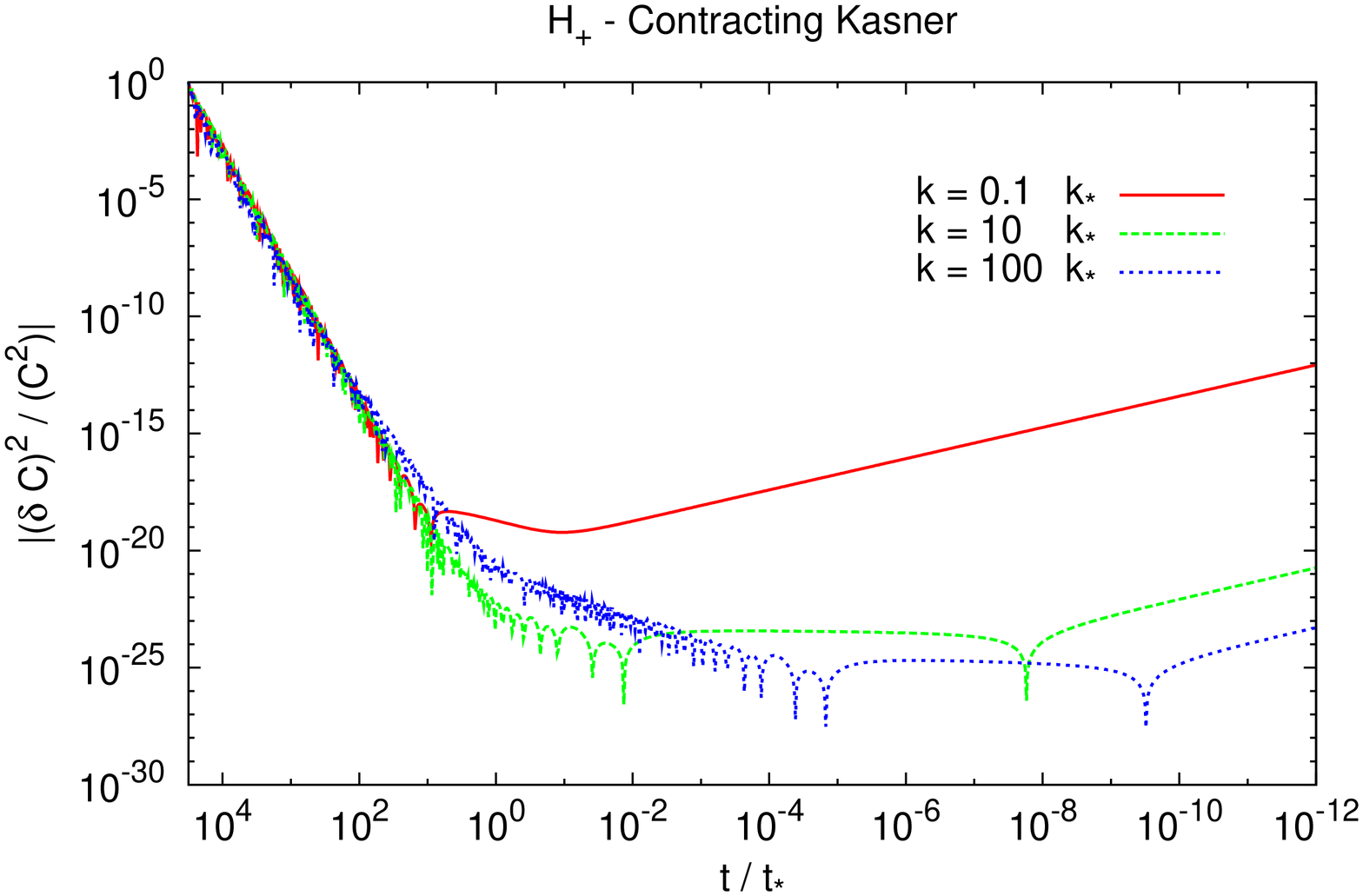}
}
\caption{
Contribution to $\delta C^2 / C^2$, defined in eq.~(\ref{pweyl}), from $H_\times$ (left panel) and 
$H_+$ (right panel) modes with some fixed comoving momentum. The times and momenta are chosen as in Figure \ref{Weyl-exp}. While in Figure \ref{Weyl-exp} the background geometry is expanding (from the singularity at $t = 0^+$), the background geometry is contracting (towards the singularity at $t = 0^-$) in the evolutions shown here.
}
\label{Weyl-con} 
\end{figure}

Once expressed in terms of absolute values of the time, the short and late time asymptotic expressions coincide with those of the expanding case, cf. eqs.~(\ref{esp-omX-exp}). As in the expanding case, the 
short scales regime occurs asymptotically far from the singularity, while the large scales regime occurs asymptotically close to the singularity (notice, however, that a mode evolves from the large scales to the short scales regime in the expanding Kasner, while from the short scales to the large scales regime in the contracting Kasner background).

Consequently, also the short and large scales asymptotical solutions are identical, once expressed in terms of $\eta$ and $\eta_* \,$.
\begin{equation}
H_\times \simeq
\left\{ \begin{array}{ll}
\frac{C_1}{\sqrt{-\eta}} \, {\rm e}^{i \frac{k_L \, 
\left( - \eta \right)^2}{2 \, \left(  - \eta_* \right)}} 
+ \frac{C_2}{\sqrt{-\eta}} \, {\rm e}^{- i \frac{k_L \, \left( - \eta \right)^2}{2 \,
\left(  - \eta_* \right)}} & \,,{\rm ~short~scales}\\
{\bar C}_1 \, \sqrt{- \eta} \, J_3 \left( 2 \, k_T \, \sqrt{ \left( - \eta_* \right) \left( -  \eta \right)} \right)
+ {\bar C}_2 \, \sqrt{- \eta} \, Y_3 \left( 2 \, k_T \, \sqrt{ \left( - \eta_* \right) \left( -  \eta \right)} \right)
& \,,{\rm ~large~scales}\\
\end{array}\right.\,.
\label{esp-hX-con}
\end{equation}
However, in contrast to the expanding case, we now keep both solutions for $H_\times$ in the large scales  regime~\footnote{We recall that this mode was disregarded in the solution 
(\ref{esp-hX-exp}), since it is a decreasing mode in that case. In the contracting case instead this mode dominates at late times.}.

For the non-oscillatory part of the Weyl tensor we find
\begin{equation}
\delta C^2  \simeq
\left\{ \begin{array}{ll}
\left( C_1 \, C_2^* \, {\rm e}^{\frac{i \, k_L \, 
\left( - \eta \right)^2}{\left( - \eta_* \right) }}
+ {\rm h.c.}  \right) \,
\frac{32 \, k_L^4}{M_p^2 \, a_*^6 \, \left( - \eta_* \right)}
& \,,{\rm ~short~scales}\\
\frac{480 \, \vert {\bar C}_2 \vert^2}{M_p^2 \, \pi^2 \, a_*^6
\, k_T^6 \, \left( - \eta \right)^9} 
& \,,{\rm ~large~scales}\\
\end{array}\right.\,.
\label{W-hX-con}
\end{equation}

Notice that the short scales asymptotic evolution agrees with the corresponding one in the expanding case. However, in the large scale regime $\delta C^2$ is now controlled by the mode that was disregarded in the expanding background.

The asymptotic behaviors (\ref{W-hX-con}) are confirmed by the fully numerical evolutions shown in
the left panel of Figure \ref{Weyl-con}. For comparison we also plot the  evolutions of the
$H_+$ mode considered in the Appendix A(b).

\subsection{Comparison with (in)stability analysis of BKL}

How does the instability of gravitational waves which we demonstrated above  fit with  the classical picture of the
universality of the rule of alternation of the Kasner epochs during contraction towards a  singularity?

Let us briefly recall  this picture.
One of the points of  the   original paper \cite{LK63} was to extend the anisotropic homogeneous contracting
Kasner solution 
to a  class of generalized Kasner solutions, describing more general inhomogeneous anisotropic geometries. It was implicitly  assumed that there are large-scale, super-horizon  inhomogeneities at  the scales exceeding
the (average) Hubble radius. 
Generic anisotropic solution shall  contain eight physically different arbitrary functions of the spatial coordinates.
However, it was identified by \cite{LK63} that the homogeneous Kasner  solution is unstable against
a particular  inhomogeneous mode for which $\vec l \times rot \, \vec l=0$ (here $\vec l$ is the Kasner axes corresponding to the direction that is expanding). In other words, this mode is unstable and is growing with time. Therefore the  monotonic (stable) inhomogeneous solution can have  only seven arbitrary functions of the spatial coordinates.
Homogeneous Kasner contraction (in Bianchi models with spatial curvature) occurs in the 
  stochastic regime of alternation  of Kasner exponents \cite{BKL70}.
This influenced the  philosophy of inhomogeneous generalization of the Kasner contraction.
Following 
\cite{BKL82}, now one can allow all eight arbitrary functions to describe generalized Kasner (Bianchi I) geometry.
According to \cite{BKL82}, the backreaction of the growing mode of the spatial instability
 alters the Kasner exponents of the patch of the contracting universe in the same manner as
they were altered in the homogeneous Kasner-oscillating  universe.  Again, in this picture  the inhomogeneity scale
is larger than the ``Hubble'' patch,  so that locally, along the time geodesics, the contraction is asymptotically homogeneous
\cite{uggla}. The 
instability induced by the spatial curvature associated with the inhomogeneous growing mode, and 
the resulting rotation of the Kasner
axes were rigorously studied with the mathematical tools of the theory of dynamical systems in  \cite{Uggla:2003fp,wain}.

In the previous Section we found instabilities of both expanding and contracting Kasner geometries against gravitational waves.
Here for comparison with the BKL analysis we focus on the GW instability in the contracting universe.
In this background, both polarizations $H_\times$ and $H_+$ are unstable in the large wavelength limit. All  physical wavelengths of the GW corresponding to the momenta $k_1, k_2, k_3$ in different directions (associated with different Kasner exponents) sooner or later will leave the Hubble radius $\sim 1/H \sim t$, independently of whether they are red- or blue-shifted. The physical frequency of  the two modes has the structure  $\omega^2 \sim \frac{k_1^2}{a^2}+\frac{k_2^2}{b^2}+\frac{k_3^2}{c^2} + f \left( H_a ,\, H_b \right)$, where the function $f$ is of order of $H_{a,b}^2$.
 For small $t$ this function dominates over the momenta terms; this turns the time dependence of the GW amplitude from the oscillating to the non-oscillating regime. The amplitude of the non-oscillating GW increases with time in this long-wavelength regime, signaling the instability which we described in the previous Section. It turns out (See Appendix \ref{appB}) that these unstable GW  modes
in the long-wavelength limit exactly coincide with the unstable solution $\vec l \times rot \, \vec l=0$  found in \cite{LK63}. Phrased in another way, the unstable solution of \cite{LK63} which destroys the monotonicity of the homogeneous Kasner contraction is nothing but 
the GW polarizations that are evolving with time
into the long-wavelength regime. This provides us with a new insight into the origin of the generalized 
inhomogeneous contracting Kasner  solution: small short-wavelength GW will eventually be stretched to become the long-wavelength inhomogeneous modes, which are unstable.

Equipped with the BKL conjecture, one  may think that for the contracting universe
 the growth of the unstable GW modes results in an
alternation of the Kasner exponents. While this conjecture is  supported
  for a contracting universe  \cite{BKL82,Garfinkle:2003bb},
it is not clear how the GW instability evolves for the expanding Kasner background.
Indeed, for contracting Kasner universe GW  leave the Hubble radius and become long-wavelength inhomogeneities, and their backreaction can be described by the impact of  the spatial curvature at the ``local'' time evolution of $a(t), b(t), c(t)$. For an expanding universe, initially long-wavelength GW enter the Hubble radius. Their backreaction can be treated with the pseudo-tensor
 $T_{\mu\nu}=M_p^2<h^{\rho\sigma}_{;\mu}h_{\rho\sigma;\nu}>$ for the high-frequency GW. 

Another interesting issue is how our unstable modes correspond to  the perturbatively small variations of the Kasner exponents
(giving contribution to the diagonal metric fluctuations) as well as small variations of the Kasner
axes (contribution to off-diagonal metric fluctuations).

Finally, one of the most interesting application of the effect is  related to
zero vacuum fluctuations of gravitons  in contracting Kasner geometry, which are  unstable.  They grow and become large scale classical GW inhomogeneities described by the random gaussian field. One may think that  the Kasner axes will be altered and 
rotated differently in different spatial domains of that random field.

We will expand these considerations in a forthcoming investigation.

\section{$\Delta T/T$ due to the residual classical GW from pre-inflation}\label{sec:obs}

The growth of the decoupled tensor polarization at large scales can leave an imprint in the amplitude and in the polarization of the CMB anisotropies. A characteristic signature is a non-diagonal correlation between different multipoles in the expansion of the anisotropies, due to initial background anisotropy.
Such extended phenomenological study is beyond the goals of the present work. However, we want to obtain a crude estimate on the limits that such a study would impose on the model, namely on the physical wavelength beyond which the statistics of the modes is anisotropic, and on the initial amplitude of the GW signal. For this reason, we compute the contribution of the GW mode to the quadrupole of the temperature anisotropies, and we impose that it does not exceed the  WMAP value 
$C_2 \simeq 3 \cdot 10^{-11} \,$ \cite{Hinshaw:2006ia}. 

As we show in appendix \ref{appC}, the $C_\ell$ coefficients of the temperature anisotropies are
related to the primordial spectrum of the GW by
\begin{eqnarray}
C_\ell = \frac{9\,\pi^3}{8}\,\frac{(\ell+2)!}{(\ell-2)!}\, \int_0^\infty \frac{d(k\,\eta_0)}{(k\,\eta_0)}\,I^2_\ell(k\,\eta_0)\int_{-1}^{+1} \frac{d\xi}{2}\,\int_0^{2\pi}\frac{d\phi_k}{2\,\pi}\left[P_{H_+} \left( {\bf k} \right) 
+ P_{H_\times} \left( {\bf k} \right) \right] \,,
\label{c-ell}
\end{eqnarray}
where $I_\ell$ is a ``window function'', which for a matter dominated universe~\footnote{With this assumption we disregard the modes that reentered the horizon during the radiation stage, which do not contribute significantly to lowest $C_\ell$'s, and  the period of late accelerated expansion; this is consistent with the present approximate computation.} is given in Eq.~(\ref{iell}). The analytic expression for $I_2 \left( k \, \eta_0 \right)$ can be found in \cite{Starobinsky:1985ww}. In this case,
the function peaks at $k \, \eta_0 \simeq 3 \,$, while it goes to zero as $\sim \left( k \, \eta_0 \right)^2$ at small argument, and as $\sim \cos \left( k \, \eta_0 \right) / \left( k \, \eta_0 \right)^2$ at large argument. The quantity $\eta_0$ is the present conformal time. For a matter dominated universe
\begin{equation}
H = \frac{2}{a_0 \, \eta} \;\; \Rightarrow \;\; \eta_0 = \frac{2}{a_0 \, H_0}
\label{md}
\end{equation}
where here $a_0$ denotes the present value of the scale factor. The comoving momentum $k$ of a mode is related to its present wavelength $\lambda$ by $\lambda = 2 \, \pi / \left( k / a_0 \right)$. Therefore
the quantity
\begin{equation}
k \, \eta_0 = 2 \, \pi \left( \frac{2 \, H_0^{-1}}{\lambda} \right)
\label{ketalambda}
\end{equation}
gives the present ratio between the size of the (particle) horizon and of the mode with comoving momentum $k \,$. As shown in Figure \ref{powertcl}, the power spectrum of the decoupled tensor polarization experiences a growth only for $k < k_{\rm iso} \,$. If the anisotropic stage is followed by a prolonged inflationary stage, these large scale modes are inflated to scales much larger than the present horizon size, $k_{\rm iso} \, \eta_0 \propto H_0^{-1} / \lambda_{\rm iso} \ll 1 \,$, and the window function $I_\ell$ suppresses the contributions of these modes to the $C_\ell$ coefficients. 

 We now relate $k_{\rm iso} \, \eta_0$ to the duration of inflation through the number of e-folds 
$N \left( k \right)$ between the moment in which the mode with comoving momentum $k$ leaves the horizon during inflation, and the end of inflation. This quantity is  \cite{Liddle:2000cg}
\begin{equation}
N \left( k \right) = 62 - {\rm ln } \left( \frac{k}{a_0 \, H_0} \right) - 
\ln \left(\frac{10^{16}GeV}{V_0^{1/4}} \right)+\Delta \ ,
\end{equation}
For a matter dominated universe $a_0 \, H_0 = 2 / \eta_0$, cf. eq. (\ref{md}); $V_0$ is the energy density in the universe at the moment in which that mode left the horizon (for the potential we are discussing, cf. Eq.~(\ref{poten}), this quantity is nearly independent of k); finally, the quantity $\Delta$ is sensitive to the details of reheating (see \cite{Podolsky:2005bw} for a discussion). Imposing that all the modes within the horizon today were sub-horizon at the beginning of inflation gives the minimal number of e-folds given in Eq.~(\ref{efold}). We define 
\begin{equation}
\Delta N \equiv N \left( k_{\rm iso} \right) - N_{\rm min} = - {\rm ln } \left( \frac{k_{\rm iso}}{a_0 \, H_0} \right)
= - {\rm ln } \left( \frac{k_{\rm iso} \, \eta_0}{2} \right)
\label{dn}
\end{equation}
Since $k_{\rm iso}$ is the comoving momentum of the modes leaving the horizon when the universe becomes isotropic, $\Delta N$ is the difference in e-folds between the duration of the isotropic inflationary stage and the minimal duration (\ref{efold}). As we discussed, a large $\Delta N$ results in a suppressed effect of the amplified GW modes on the observed CMB anisotropies.

We compute only the contribution of $H_\times$ to the $C_2$ coefficient in Eq.~(\ref{c-ell}). Figures~\ref{powertcl}
 and \ref{powerxi} give the power spectrum of this mode at the end of inflation normalized by the initial power spectrum $P_0$. This quantity is related to the initial condition at the start of the anisotropic stage, and it was left unspecified. For definiteness, we parametrize it with an overall amplitude and with a  power law dependence on the scale
\begin{equation}
P_0 = {\cal A}_0 \times \left( \eta_0 \, k \right)^n \ ,
\label{defAn}
\end{equation}
where ${\cal A}_0$ is the initial power at a scale which parametrically corresponds to the present horizon size. We numerically perform the angular integral in (\ref{c-ell}), to find
\begin{equation}
{\bar P} \left( \frac{k}{k_{\rm iso}} \right) \equiv
\int_{-1}^1 d \xi P_\times \left( k ,\, \xi \right) \simeq 2\times10^3 \, \frac{\vert h_0 \vert}{\sqrt{V_0}/\phi_0} \, 
{\cal A}_0 \, \left( \eta_0 \, k \right)^n \times   
\left\{ \begin{array}{ll}
1 & , \;\; k \lta k_{\rm iso} \\
\left( \frac{k_{\rm iso}}{k} \right)^3 & , \;\; k \gta k_{\rm iso}
\end{array} \right.
\label{eqpbar}
\end{equation}
where the proportionality to $\vert h_0 \vert$ is related to the growth of the power spectrum during the anisotropic era, as explained at the end of Section \ref{sec:decoupled}. The dependence on $k$ is due to the fact that the angular integral is dominated by modes of small angle ($k_L \gg k_T$), cf. Figure \ref{powerxi}, where the growth of the spectrum is flat for $k \lta k_{\rm iso}$, and scales as $\sim k^{-3}$ at larger $k \,$.

\begin{figure}[h]
\centerline{
\includegraphics[width=0.6\textwidth,angle=270]{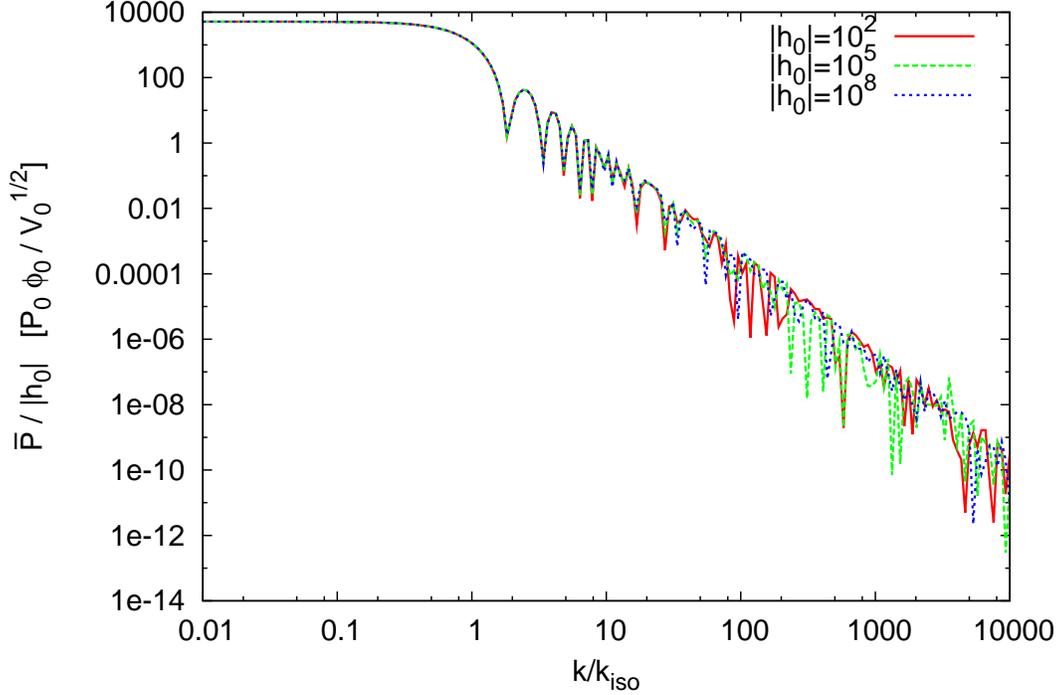}
}
\caption{Angular integral of the power spectrum, defined in Eq.~(\ref{eqpbar}). We show the result 
normalized by the initial power spectrum, starting from different values of $\vert h_0 \vert$ (given in units of $\sqrt{V_0} / \phi_0$; larger values of this quantity corresponds to a longer anisotropic stage), and rescaling it by $\vert h_0 \vert$, to explicitly show the linear dependence of the power spectrum on this quantity.}
\label{barP} 
\end{figure}

The function ${\bar P}$ is shown in Figure \ref{barP}, normalized by $\vert h_0 \vert$ and by the initial power spectrum $P_0 \,$. The dependences on $\vert h_0 \vert$ and on $k$ 
given in eq. (\ref{eqpbar}) are manifest in the Figure.

We can now perform the final integral over momenta in (\ref{c-ell}). Taking into account the large and small argument dependence of $I_2$  and that of ${\bar P}$ 
given in eq. (\ref{eqpbar}), we see that the integrand behaves as $\sim k^{n-3}$ at small $k$, and as $\sim k^{n-8}$ at large $k$. Therefore, the integral converges for a wide range for the initial slope defined in Eq.~(\ref{defAn}), namely $-4 < n < 7 \,$.

\begin{figure}[h]
\centerline{
\includegraphics[width=0.6\textwidth,angle=270]{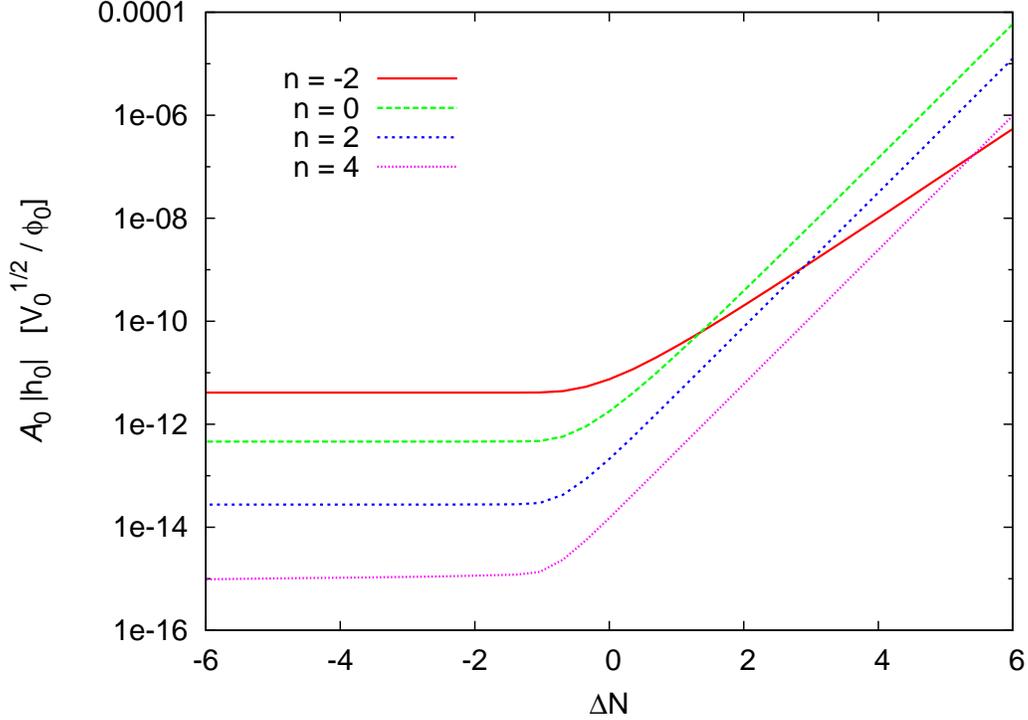}
}
\caption{Limit on the initial amplitude of the power spectrum, times its growth during the anisotropic stage. The quantity $\Delta N$, defined in eq. (\ref{dn}), is the number of e-folds of isotropic inflation minus the minimum usually required to homogenize the universe, cf. eq. (\ref{efold}). For a longer duration of the inflationary stage, the modes experiencing the growth are inflated to larger scales than the present horizon, and the resulting effect is suppressed (weaker limit). ${\cal A}_0$ and $n$, are, respectively, the amplitude and the slope of the initial power spectrum, see eq. (\ref{defAn}), while $\vert h_0 \vert$ is proportional to the duration of the anisotropic stage.}
\label{final-lim} 
\end{figure}

After performing the integral, we impose that the resulting $C_2$ does not exceed the  WMAP value $C_2 \simeq 3 \cdot 10^{-11} \,$ \cite{Hinshaw:2006ia}. This results in an upper limit on ${\cal A}_0 \, \vert h_0 \vert$ (the initial overall amplitude, times the growth during the anisotropic phase) for any value of $k_{\rm iso}$ and $n \,$. We show this in Figure~\ref{final-lim}. As explained after Eq.~(\ref{ketalambda}), the limit weakens at small values of $k_{\rm iso} \, \eta_0$, corresponding to 
a longer duration of the inflationary stage.

\begin{figure}[h]
\centerline{
\includegraphics[width=0.6\textwidth,angle=270]{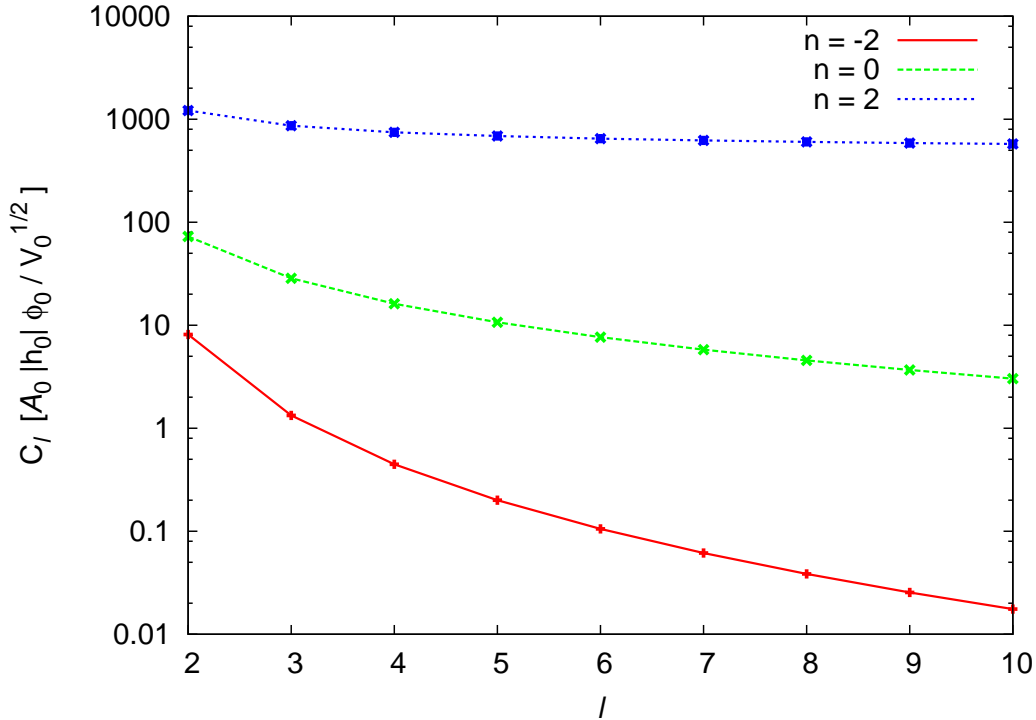}
}
\caption{First $C_\ell$ coefficients generated by the GW, normalized by the initial amplitude and the duration of
 the anisotropic era (controlled by $\vert h_0 \vert$). The tree curves corresponds to three different slopes of the initial power in the GW, eq. (\ref{defAn}).}
\label{Cl} 
\end{figure}

Finally, we show in Figure~\ref{Cl} the first few $C_\ell$'s, normalized by the initial amplitude of the GW and the length of the anisotropic era (controlled by $\vert h_0 \vert$), for a few values of the slope (\ref{defAn}).
As expected, for moderate slopes, the spectrum of $C_\ell$ decreases with $\ell \,$. This is due to the fact that larger angular scales (small $\ell$) are affected by the modes with lower momenta, which grow more during the anisotropic stage. The three spectra shown in the figure can be fitted by a single power-law (with an accuracy up to about $10 \%$):
\begin{equation}
C_\ell \simeq \left\{ \begin{array}{l}
80 \, \ell^{-3.8} \;\;\;,\;\;\; n = -2 \\
248 \, \ell^{-1.9} \;\;\;,\;\;\; n = 0 \\
1430 \, \ell^{-0.4} \;\;\;,\;\;\; n = 2
\end{array} \right.
\end{equation}

We see that the angular spectrum roughly decreases with $\ell$ as $C_\ell \sim 1/ \ell^p$, where the exponent $p$ depends on the  power-spectrum of the  classical GW at the initial hypersurface.

\section{Discussion}\label{sec:sum}

We found a new effect of instability of the gravitational waves in an expanding and contracting
Kasner geometries. We demonstrated the effect for a particular choice of the Kasner exponents
$(-1/3, 2/3, 2/3)$, but we expect it is rather generic. This particular choice allows to simplify
the description of GW in such a way that their wave equations is manifestly depending on the
momenta. For the contracting Kasner geometry,  we found that   our unstable  GW mode
is identical to the unstable large-scale inhomogeneous mode first identified in  \cite{LK63}-\cite{BKL82}
 (for arbitrary Kasner exponents ($p_1,p_2, p_3$)). Backreaction of this  unstable mode is conjectured to
alter the Kasner exponents, differently in different spatial patches, depending on the spatial profile
of the initial GW.

The Kasner geometry is a rather universal asymptotic solution for many interesting situations: it describes approach towards  black holes or cosmological singularities
 (including higher dimensional and supersymmetric cases \cite{Damour:2000hv}), and it describes generic anisotropic expansion from singularity prior to inflation.

All of these situations are inevitably accompanied by quantum fluctuations of the gravitons), and, possibly, also by  classical gravitational waves. 
There is a long list of questions arising in connection with a new effect of the gravity waves instability
in anisotropic geometry. 
We have to investigate how the instability growth depends on the Kasner exponents $p_1,p_2, p_3$.
It will be interesting to understand if the instability of the classical GW  also results in the instability
of the graviton zero vacuum fluctuations.
It will also be worth to understand the impact of the GW instability on the structure of 
the singularity inside the black hole and cosmological singularity.
It is also interesting to investigate the backreaction of the GW instability on the contracting and expanding anisotropic geometries depending on the initial GW profile.

In this paper we considered the effect of GW instability in the context of the
anisotropic pre-inflationary stage. The
transition from Kasner expansion to inflation terminates the effect of GW instability
but leaves  classical GW signal as the initial conditions for inflation.
If inflation does not last  very long, then the residual GW can contribute to the
CMB temperature anisotropies. Since GW polarizations and the scalar mode of cosmological fluctuations
behave differently during anisotropic pre-inflation,
we can consider only the leading  contribution, namely, $H_\times$ mode of the GW polarization.
In this paper we calculated its contribution to the total $\Delta T/T$ anisotropy angular power spectrum.
The angular power spectrum of the signal decreases with $\ell$ in a power-law manner and depends
on the initial spectrum of the classical GW fluctuations,  $C_\ell \sim 1/ \ell^p$, where the exponent
$p$ depends on the  power-spectrum of the  classical GW at the initial hypersurface.
It is interesting to note that this result is qualitatively similar to the result of \cite{linde08},
where the impact of GW from the our-universe-bubble nucleation was considered.

The  signal from $H_\times$ is rather anisotropic, and there is an interesting question about the anisotropy of its multipole structure $\langle a_{\ell m} \, a_{\ell' m'}^* \rangle$, which is intriguing in connection with an apparent alignment of the low multipoles of $\Delta T/T$.  While such an analysis is beyond the aims of the
present work, we have estimated the initial conditions (initial amplitude of the GW, versus the duration of inflation) which can lead to potentially observable effects.
We leave a more extended analysis to future investigation.

\vspace{1cm}

{\bf \large Acknowledgements}

\bigskip

\noindent We thank J.R.~Bond, C.R.~Contaldi, T.~Damour,  I.~Khalatnikov,  A.~Linde, C.~Pitrou, M.~Sasaki, 
A.~Starobinsky,  J.P.~Uzan and J.~Weinwright
 for useful discussions and correspondence. The work of A.E.G. and M.P. was partially supported by  the DOE 
grant DE-FG02-94ER-40823. LK was supported by NSERC and CIFAR.


\appendix

\section{Contribution of the $H_+$ mode to the Weyl \\ invariant} \label{appA}

In this Appendix we compute the contribution of the $H_+$ mode to the square of the Weyl tensor for a pure Kasner background. We treat the expanding and contracting cases separately. This presentation follows the analogous one for the $H_\times$ mode placed in Subsection \ref{subHX} of the main text.

\subsubsection{ $H_+$ mode during expansion}

Substituting (\ref{abeta2}) into (\ref{omega}) we have the large and short scale expansions
\begin{equation}
\omega_+^2 \simeq
\left\{ \begin{array}{ll}
\frac{1}{4 \, \eta^2} +  k_T^2 \, \frac{\eta_*}{\eta} & \,,{\rm ~large~scales}\\
k_L^2 \, \left( \frac{\eta}{\eta_*} \right)^2 & \,,{\rm ~short~scales}\\
\end{array}\right.\,.
\label{esp-omp-exp}
\end{equation}

Consequently, we have the asymptotic solutions
\begin{equation}
H_+ \simeq
\left\{ \begin{array}{ll}
C_1 \, \sqrt{\eta} \, J_0 \left( 2 \,  k_T \, \sqrt{\eta_* \, \eta} \right)
& \,,{\rm ~large~scales}\\
\frac{{\bar C}_1}{\sqrt{\eta}} \, {\rm e}^{i \frac{k_L \, \eta^2}{2 \, \eta_*}} + \frac{{\bar C}_2}
{\sqrt{\eta}} \, {\rm e}^{-i \frac{k_L \, \eta^2}{2 \, \eta_*}}
& \,,{\rm ~short~scales}\\
\end{array}\right.\,.
\label{esp-hp-exp}
\end{equation}
where again in the early time solution we have disregarded a decaying mode that would diverge at 
the initial singularity. Notice that the late time expression is the same as for the $H_\times$ mode (since the late time expansions of $\omega_\times^2$ and $\omega_+^2$ coincide).

For the linear part of the Weyl tensor we find
\begin{equation}
C \, \delta C \simeq {\rm e}^{-i k_L \, x - i k_T \, y} \times
\left\{ \begin{array}{ll}
- \frac{36 \, \sqrt{2} \, C_1 \, \eta_*^{5/2}}{M_p \, a_*^5 \, \eta^6}
& \,,{\rm ~large~scales}\\
\left(  {\bar C}_1 \, {\rm e}^{\frac{i \, k_L \, \eta^2}{2 \, \eta_*}}
+ {\bar C}_2 \, {\rm e}^{- \frac{ i \, k_L \, \eta^2}{2 \, \eta_*}}
\right) \frac{12 \, \sqrt{2} \, k_T^2 \, \eta_*^{7/2}}{M_p \, a_*^5 \, \eta^6}
& \,,{\rm ~short~scales}\\
\end{array}\right\}
+ {\rm h.c.}
\label{Wlin-hp-exp}
\end{equation}

The non-oscillatory part of the $2$nd order term in the Weyl square is
\begin{equation}
\delta C^2  \simeq
\left\{ \begin{array}{ll}
\frac{324   \, \vert C_1 \vert^2 \, \eta_*^3}{
M_p^2 \, a_*^6 \, \eta^6}
& \,,{\rm ~large~scales}\\
\left( {\bar C}_1 \, {\bar C}_2^* \, {\rm e}^{\frac{i \, k_L \, \eta^2}{\eta_*}}
+ {\rm h.c.}  \right) \,
\frac{64 \, k_L^6 \, \eta^2}{M_p^2 \, a_*^6 \, k_T^4 \, \eta_*^5}
& \,,{\rm ~short~scales}\\
\end{array}\right.\,.
\label{W-hp-exp}
\end{equation}

This behavior is confirmed by the fully numerical solutions shown in the right panel of 
Figure~\ref{Weyl-exp}. It is interesting to compare the contribution of the $H_\times$ mode and that of the $H_+$ mode. In the large scale regime only the contribution of the $H_\times$ mode grows with respect to the background. This is consistent with the amplification of this polarization in the Kasner era that we have found in Section \ref{sec:decoupled}. In the small scales regime, the contribution of both modes is instead growing with respect to the background. Although the two modes evolve identically in this regime (since they satisfy the same asymptotic equation), the time evolution of the corresponding 
$\delta C^2$ term is different, since the two modes enter differently in the Weyl tensor.

\subsubsection{ $H_+$ mode during contraction}

Plugging (\ref{abeta2}) into (\ref{omega}) we have the early / late 
time expansions
\begin{equation}
\omega_+^2 \simeq
\left\{ \begin{array}{ll}
k_L^2 \, \left( \frac{-\eta}{-\eta_*} \right)^2 
& \,,{\rm ~short~scales}\\
\frac{1}{4 \, \left( - \eta \right)^2} + k_T^2 \, \frac{-\eta_*}{-\eta} 
& \,,{\rm ~large~scales}\\
\end{array}\right.\,.
\label{esp-omp-con}
\end{equation}

The similarities and differences between the $H_+$ mode in the expanding and contracting cases are identical to those discussed in the previous Subsection for the $H_\times$ mode. We have the asymptotic solutions
\begin{equation}
H_+ \simeq
\left\{ \begin{array}{ll}
\frac{ C_1}{\sqrt{- \eta}} \, {\rm e}^{i \frac{k_L \,  \left( - \eta \right)^2}{2 \, \left( - \eta_* \right)}} 
+ \frac{ C_2}{\sqrt{- \eta}} \, {\rm e}^{-i \frac{k_L \, \left( - \eta \right)^2}{2 \, \left( - \eta_* \right)}}
& \,,{\rm ~short~scales}\\
{\bar C}_1 \, \sqrt{-\eta} \, J_0 \left( 2 \, k_T \, \sqrt{\left( - \eta_* \right) \left( - \eta \right)} \right)
+ {\bar C}_2 \, \sqrt{-\eta} \, Y_0 \left( 2 \, k_T \, \sqrt{\left( - \eta_* \right) \left( - \eta \right)} \right)
& \,,{\rm ~large~scales}\\
\end{array}\right.\,.
\label{esp-hp-con}
\end{equation}
This leads to the following asymptotic evolution for the linear term in the square of the Weyl tensor
\begin{equation}
C \, \delta C \simeq {\rm e}^{-i k_L \, x - i k_T \, y} \times
\left\{ \begin{array}{ll}
\left(   C_1 \, {\rm e}^{\frac{i \, k_L \, \left( - \eta \right)^2}{2 \, \left(  - \eta_* \right)}}
+  C_2 \, {\rm e}^{- \frac{ i \, k_L \, \left( - \eta \right)^2}{2 \, \left( - \eta_* \right)}}
\right) \frac{12 \, \sqrt{2} \, k_T^2 \, \left( - \eta_* \right)^{7/2}}{M_p \, a_*^5 \, 
\left(  - \eta \right)^6}
& \,,{\rm ~short~scales}\\
- \left[ 3 \, \pi \, {\bar C}_1 + 6 \, {\bar C}_2 \left(
{\rm ln } \left( k_T \, \sqrt{\left( - \eta_* \right) \left( - \eta \right)}
\right)  + \gamma - \frac{2}{3} \right) \right]  \\
\quad\quad\quad\quad
\times   \frac{ 12 \, \sqrt{2} \left( - \eta_* \right)^{5/2}}{\pi \, M_p \, a_*^5 \,
\left( - \eta \right)^6}
& \,,{\rm ~large~scales}\\
\end{array}\right\}
+ {\rm h.c.}
\label{Wlin-hp-con}
\end{equation}
and for the non-oscillatory part in the quadratic term of the square of the Weyl tensor
\begin{equation}
\delta C^2  \simeq
\left\{ \begin{array}{ll}
\left(  C_1 \,  C_2^* \, {\rm e}^{\frac{i \, k_L \, \eta^2}{\eta_*}}
+ {\rm h.c.}  \right) \,
\frac{64 \, k_L^6 \, \left( - \eta \right)^2}{M_p^2 \, a_*^6 \, k_T^4 \, \left( - \eta_* \right)^5}
& \,,{\rm ~short~scales}\\
\frac{-120   \, \vert {\bar C}_2 \vert^2 \, \left( - \eta_* \right)^2}{
M_p^2   \, \pi^2 \, a_*^6 \, k_T^2 \, \left( - \eta \right)^7}
& \,,{\rm ~large~scales}\\
\end{array}\right.\,.
\label{W-hp-con}
\end{equation}

This behavior is confirmed by the fully numerical evolutions shown in the right panel 
of Figure \ref{Weyl-con}.

\section{Instability of Kasner Solution: KL vs GW} \label{appB}

Our analysis of the squared Weyl invariant shows that the Kasner background is unstable due to the amplification of the GW perturbations $H_+$ and $H_\times$. This is manifest by the growth of $\delta C / C$ and $\delta C^2 / C^2$ summarized in table \ref{tab1}. For the contracting background we found
that the contribution to $\delta C^2 / C^2$ of the polarizations $H_+$ and $H_\times$ grow, respectively, as $\vert t \vert^{-2/3}$ and as $\vert t \vert^{-2}$ asymptotically close to the singularity. Conversely, in the expanding case, the Weyl invariant is regular near the singularity. The reason for this discrepancy is that, in the expanding case, we disregarded the Neumann term $Y_0$ (for $H_+$) and $Y_3$ (for $H_\times$) in the solutions of the evolution equations (\ref{eqomtomp}) for the two modes. However these two modes are generally excited as the contracting Kasner background solution approaches the singularity.

The purpose of this Appendix is to show explicitly that this instability is not inconsistent with the lore of alternation of  Kasner exponents \cite{BKL70}. On the contrary, the instability in the contracting case was pointed out already in \cite{LK63}, where it was shown that the Kasner background is stable provided one physical condition is imposed. As we now show, the physical condition imposed by \cite{LK63} precisely eliminates the unstable $Y_0$ and $Y_3$ solutions.

We start by quoting the results of \cite{LK63} where the authors considered a perturbed vacuum Kasner solution in synchronous gauge and did a stability analysis near the singularity. In the main text, we describe the contracting Kasner background using negative time. In \cite{LK63}, instead, positive time was used, and the approach to the singularity was studied in the $t \rightarrow 0^+$ limit. In this Appendix, we adopt the time convention of \cite{LK63}. Moreover, we specify the analysis of \cite{LK63} to the axisymmetric  $p_1 = -1/3$, $p_2=p_3 = 2/3$ case. For this choice, the solution obtained in 
\cite{LK63} for $t\rightarrow 0^+$ reads
\begin{eqnarray}
h^1_1 &=& A_1 + B_ 1\,\ln t \quad\,,\quad
h^2_2 = A_2 + B_2 \, \ln t \quad\,,\quad h^3_3 = A_3 + B_3 \, \ln t \,,\nonumber\\
h_{12} &=& t^{-2/3} \left[C_{12}- \frac{9}{32}\,k_3\,\left(k_2\,C_{13} - k_3\,C_{12}\right) t^{2/3} \right]+t^{4/3}\,C_{21} \,, \nonumber\\
h_{13} &=& t^{-2/3} \left[C_{13}- \frac{9}{32}\,k_2\,\left(k_3\,C_{12} - k_2\,C_{13}\right) t^{2/3} \right]+t^{4/3}\,C_{31} \,, \nonumber\\
h_{23} &=& t^{4/3} \left(C_{23} +C_{32}\right)\,,
\label{blsol}
\end{eqnarray}
with constants $A_i$,$B_i$, $C_{ij}$ satisfying the constraints
\begin{eqnarray}
&&B_1 +B_2 +B_3 = -\frac{1}{3}B_1+ \frac{2}{3} \left(B_2+B_3\right)=0\,,\nonumber\\
&&k_1\left(B_1-A_2-A_3\right) +2\,\left(k_2\,C_{21} +k_3\,C_{31} \right) = 0\,,\nonumber\\
&&k_2 \left(B_2 + A_1\right) -2\,k_1\,C_{12} = 0 \,,\nonumber\\
&&k_3 \left(B_3 + A_1\right) -2\,k_1\,C_{13} = 0\,.
\label{constraints}
\end{eqnarray}
As a result of the synchronous gauge, the solutions (\ref{blsol}) still have freedom due to gauge artifacts; the corresponding gauge contributions to the perturbations are given in eqs (F3) of \cite{LK63}. Using this freedom, the Authors of \cite{LK63} set $C_{12} = C_{31} = C_{23}+C_{32}=0$. \footnote{They are in fact setting $C_{23}=0$ in the general case. However, the time dependences of both terms in $h_{23}$ are the equal to each other for the axisymmetric $p_2=p_3$ case; therefore, in this case, the removal of the gauge artifact corresponds to $C_{23}+C_{32}=0$.} After this gauge fixing, one sees that the mode $h_{13}$ diverges at the singularity. For this reason,  ref. \cite{LK63} concluded that the Kasner background is stable only provided the condition $C_{13}=0$ is imposed. We stress that setting $C_{13}$ is not a gauge choice, but rather the suppression of a physical degree of freedom. This physical constraint, along with the relations (\ref{constraints}), implies that $B_2=0$.

We now turn back to our analysis. For $t \rightarrow 0^+$, eqs.~(\ref{eqomtomp}) admit the solutions
\begin{eqnarray}
H_+ &=& t^{1/3}\,\left[D_1\,J_0 (3\,k_T\,t^{1/3}) +D_2\,Y_0 (3\,k_T\,t^{1/3}) \right]\,,\nonumber\\
H_{\times} &=& t^{1/3}\,\left[E_1\,J_3 (3\,k_T\,t^{1/3}) +E_2\,Y_3 (3\,k_T\,t^{1/3}) \right]\,.
\label{earlysols}
\end{eqnarray}
The modes $J_0$ and $J_3$ are regular at the singularity, whereas $Y_0$ and $Y_3$ diverge, thus resulting in the instability. We now show that the conditions $C_{12} = C_{13} = B_2$ in \cite{LK63} correspond to having $D_2 = E_2=0$ in our case.  To do such a comparison, we need to change from   the gauge chosen in the main text to the synchronous gauge adopted in \cite{LK63}. We do so through a general infinitesimal transformation $x^\mu \rightarrow x^\mu+\xi^\mu$ , with $\xi^\mu = \left( \xi^0 ,\, \partial_1\,\xi^1 ,\, \partial^i \xi +\xi^i \right)$ where $\xi^i$ is a transverse 2d vector. From the transformed metric $\delta g_{\mu\nu} \rightarrow \delta g_{\mu \nu} - \mathcal{L}_\xi g_{\mu\nu}^{(0)}$, we find that the parameters $\xi^\mu$ which take our gauge to the synchronous coordinates are
\begin{eqnarray}
\xi^0 &=& \frac{1}{a_{\rm av}} \int^t dt' \,\Phi(t') + \frac{1}{a_{\rm av}}\,f^0\,,\nonumber\\
\xi^1 &=& \int^t dt'\,\frac{\chi(t')}{a(t')} + \int^t \frac{dt'}{a^2(t')} \int^{t'} dt'' \, \Phi(t'') +f^0 \int^t \frac{dt'}{a^2(t')} +f^1\,,\nonumber\\
\xi &=& \int^t dt'\,\frac{B(t')}{b(t')} + \int^t \frac{dt'}{b^2(t')} \int^{t'} dt'' \, \Phi(t'') +f^0 \int^t \frac{dt'}{b^2(t')} +f\,,\nonumber\\
\xi^i &=& \int^t dt'\,\frac{B_i(t')}{b(t')} +f^i\,,
\end{eqnarray}
where $f^0 , \,f^1,\, f, \,f^i$ are integration constants and $k_i f^i =0$. The perturbations $\Phi$, $\chi$, $B$ and $B_i$ are the modes in the decomposition given in the main text. Using their relations to the canonical modes $H_+$ and $H_{\times}$, along with the early time solutions (\ref{earlysols}), we calculate the metric perturbations in synchronous gauge. The relevant components for the present discussion are
\begin{eqnarray}
h_{12} &=& - \frac{16\,i\,\sqrt{2} \, k_3}{27\,\pi\,k_T^4\,M_p}\,E_2 \left(t^{-2/3} + \frac{9}{8}\,k_T^2 +\frac{81}{64}\,k_T^4\,t^{2/3}\right) 
-\frac{12}{5}\,f^0\,k_1\,k_2\,t + k_1\,k_2\,f^1\,t^{-2/3}+\mathcal{O}(t^{4/3})
\,,\nonumber\\
h_{13} &=& \frac{16\,i\,\sqrt{2} \, k_2}{27\,\pi\,k_T^4\,M_p}\,E_2 \left(t^{-2/3} + \frac{9}{8}\,k_T^2 +\frac{81}{64}\,k_T^4\,t^{2/3}\right) 
-\frac{12}{5}\,f^0\,k_1\,k_3\,t + k_1\,k_3\,f^1\,t^{-2/3}+\mathcal{O}(t^{4/3})\,,\nonumber\\
h^2_2 &=& -\frac{\sqrt{2}}{3\,\pi\,M_p} \left[3\,\pi\,D_1 +2\,\left(3\,\gamma-2\right)D_2 + 6\,D_2\, \ln \left(\frac{3\,k_T}{2}\right) \right]
\nonumber\\
&&+\frac{2\,\sqrt{2}}{27\,\pi\,k_T^5\,M_p}\left[-\,32\,i\,k_1\,k_2\,\frac{k_3}{k_T} E_2 -9\,D_2\,k_T^3 \left(k_3^2-k_2^2\right) \right]\ln t
\nonumber\\
&&-2\,f_0\left(3\,k_2^2\,t^{-1/3} +\frac{2}{3\,t}\right) + 2\,k_2\left(i\,f_2 +k_2\,f\right)  + \mathcal{O}(t^{2/3})\,.
\label{gkpsol}
\end{eqnarray}
In eqs (\ref{blsol}), setting $C_{12}=0$ (gauge choice) and $C_{13}=0$ (physical choice) eliminates the terms up to $\mathcal{O}(t^{4/3})$ in $h_{12}$ and $h_{13}$. On the other hand, from our solution (\ref{gkpsol}), we see that these terms in $h_{1i}$ vanish only if $E_2 =0$. Furthermore, the choice $C_{12}=C_{13}=0$ implies that $B_2=0$, now eliminating the log term in $h^2_2$, which corresponds to $D_2=0$ in our solutions. In other words, the stability conditions derived in \cite{LK63} correspond to removing the Neumann functions in the early time solutions (\ref{earlysols}) of both $H_+$ and $H_\times$.

\section{Explicit computation of $C_{\ell \ell' m m'}$} \label{appC}

Here, we prove the relation (\ref{c-ell}) of the main text. This expression gives the $C_\ell$ coefficients of the temperature anisotropies in terms of the value of the power spectrum of GW at the end of inflation. Therefore, this computation is performed in an isotropic universe. The effects of the anisotropic stage are encoded in the power spectra that, on large scales, depend also on the orientation of the modes.

We start by decomposing the temperature perturbations in direction ${\bf \hat{p}}$ measured by an observer at ${\bf x_0}$ and at conformal time $\eta_0$, into spherical harmonics
\begin{equation}
 \frac{\delta T}{T}({\bf \hat{p}}) = \sum_{\ell,m} a_{\ell m}\,Y_{\ell m} ({\bf \hat{p}} ) \,.
\label{definedt}
\end{equation}
Assuming that the perturbations are Gaussian, their statistics are completely specified by the second order correlations, which can be written by inverting (\ref{definedt})
\begin{equation}
C_{\ell \ell' m m'} \equiv \langle a_{\ell m} a^\star_{\ell'm'}\rangle =
\int d\Omega_{\hat{p}} \, d\Omega_{\hat{p}'} \, \left\langle \frac{\delta T}{T}({\bf \hat{p}}) \;\frac{\delta T}{T}({\bf \hat{p}'}) \right\rangle \, Y^\star _{\ell m} ({\bf \hat{p}}) Y _{\ell' m'} ({\bf \hat{p}'})\,.
\label{correl1}
\end{equation}
Assuming instantaneous recombination, we can use the well-known solution to 
the first order Boltzmann equation for gravity waves:
\begin{equation}
\frac{\delta T}{T}({\bf \hat{p}}) = -\frac{1}{2}\,\int_{\eta_{\rm LSS}}^{\eta_0} d\eta\,\frac{\partial\, h_{ij}}{\partial\,\eta}\, \hat{p}^i \,\hat{p}^j\,,
\label{sachswolfe}
\end{equation}
where $h_{ij} \equiv \delta g_{ij} /a^2$. We expand the gravitational perturbations in terms of plane waves as
\begin{equation}
h_{ij} ({\bf x}\,,\eta_0) = \sum_{\lambda= +, \times} \, \int \frac{d^3k}{\left(2\,\pi\right)^{3/2}}\,{\rm e}^{i\,{\bf k}\cdot{\bf x}}\,\epsilon ^{ (\lambda)}_{ij}({\bf k})\, h^{(\lambda)} ({\bf k},\eta) \, a_{\bf k}^{(\lambda)}
\label{decomposeh}
\end{equation}
where $(\lambda)$ designates the two polarizations $(+)$ and $(\times)$, $\epsilon ^{ (\lambda)}_{ij}({\bf k})$ is the normalized polarization tensor satisfying $\epsilon ^{ (\lambda)}_{ij}({\bf k}) \epsilon _{ (\lambda')}^{ij} ({\bf k}) = \delta^{(\lambda)}_{(\lambda')}$, and $a_{\bf k}^{(\lambda)}$ are Gaussian random variables with unit dispersion, ie. $\langle a_{\bf k}^{(\lambda)}\,,a_{\bf k'\,(\lambda')}^\star\rangle = \delta^{(3)}({\bf k}-{\bf k'})\,\delta^{(\lambda)}_{(\lambda')}$. Using the solution (\ref{sachswolfe}) along with the decomposition (\ref{decomposeh}) in the expression (\ref{correl1}), we get
\begin{eqnarray}
C_{\ell \ell' m m'} &=& \frac{1}{4} \sum_\lambda \int \frac{d^3k}{(2\,\pi)^3} \int d\Omega_{\hat{p}} \, d\Omega_{\hat{p}'} 
\, Y^\star _{\ell m} ({\bf \hat{p}}) Y _{\ell' m'} ({\bf \hat{p}'})\nonumber\\
&&
\quad\quad\quad\quad\times\left[ \int d\eta \left( \frac{\partial}{\partial \eta} h^{(\lambda)} ({\bf k},\eta)\right) {\rm e}^{i\,{\bf k} \cdot {\bf \hat{p}}(\eta-\eta_0)} \epsilon ^{ (\lambda)}_{ij} ({\bf k}) \hat{p}^i \hat{p}^j \right] \nonumber\\
&&
\quad\quad\quad\quad\times\left[ \int d\tilde{\eta} \left( \frac{\partial}{\partial \tilde{\eta}} h^{(\lambda)} ({\bf k},\tilde{\eta})\right) {\rm e}^{i\,{\bf k} \cdot {\bf \hat{p}'}(\tilde{\eta}-\eta_0)} \epsilon ^{ (\lambda)}_{ab} ({\bf k}) \hat{p}^{\prime\,a} \hat{p}^{\prime\,b} \right]^\star \,,
\end{eqnarray}
where we noted that ${\bf x} = {\bf x_0} + {\bf \hat{p}}(\eta-\eta_0)$ describes the geodesics of the photons with direction ${\bf \hat{p}}$, observed at $({\bf x_0},\eta_0)$.

The contraction $\epsilon ^{ (\lambda)}_{ij} ({\bf k}) \, \hat{p}^i \, \hat{p}^j$ can be calculated in terms of the angles of ${\bf \hat{p}}$ defined with respect to a coordinate system with a $z$--axis coinciding with ${\bf \hat{k}}$, as \cite{White:1992fj}
\begin{equation}
\epsilon ^{ (\lambda)}_{ij} ({\bf k}) \, \hat{p}^i \, \hat{p}^j = \left[ \delta^{(\lambda)}_{(+)} \cos(2\,\phi_{\hat{p}_{\hat{k}}}) +\delta^{(\lambda)}_{(\times)} \sin(2\,\phi_{\hat{p}_{\hat{k}}})  \right] \sin^2 \theta_{\hat{p}_{\hat{k}}}\,.
\end{equation}
To take the angular integral over $d \Omega_{\hat{p}}$ $(d \Omega_{\hat{p}'})$, we first rotate the spherical harmonics to the angular basis $\hat{p}_{\hat{k}}$ ($\hat{p}'_{\hat{k}}$) with
\begin{equation}
Y_{\ell m} ({\bf \hat{p}}) = \sum_M D^{\ell\,\star}_{m\,M}({\bf {\hat{k}}})\,Y_{\ell M} ({\bf \hat{p}_{\hat{k}}})\,,
\end{equation}
where $D^{\ell\,\star}_{m\,M}$ are the Wigner coefficients satisfying $\sum_M D^{\ell}_{M\,m}\,D^{\ell\,\star}_{M\,m'} = \delta_{mm'}$. Finally, the angular integrals can be evaluated as \cite{White:1992fj}
\begin{eqnarray}
\int d\Omega_{\hat{p}_{\hat k}} \, Y^\star _{\ell m}({\bf \hat{p}_{\hat k}}) \, {\rm e}^{i\,{\bf k} \cdot {\bf \hat{p}}(\eta-\eta_0)} \,\epsilon ^{ (\lambda)}_{ij} ({\bf k}) \,\hat{p}^i \, \hat{p}^j  &=&
- (-i)^\ell\,\pi\,\sqrt{\frac{\left(2\,\ell+1\right)\left(\ell+2\right)!}{2\,\left(\ell-2\right)!}}
\frac{J_{\ell+1/2}\left[k\left(\eta_0-\eta\right)\right]}{\left[k\left(\eta_0-\eta\right)\right]^{5/2}}
\nonumber\\
&&\times\left[\left(\delta_{(+)} ^{(\lambda)} -i\,\delta_{(\times)} ^{(\lambda)}\right) \delta_{M,2} + \left(\delta_{(+)} ^{(\lambda)} + i\,\delta_{(\times)} ^{(\lambda)}\right) \delta_{M,-2}\right]\,.\nonumber\\
\label{integwhite}
\end{eqnarray}
Ignoring the effect on the (small scale) modes which enter the horizon during the radiation dominated era, the solutions for the gravity waves during matter domination can be written in terms of the primordial one as \cite{Starobinsky:1985ww}, \cite{Turner:1993vb}
\begin{equation}
h^{(\lambda)} ({\bf k},\eta) = 3\,\sqrt{\frac{\pi}{2}} \,\frac{J_{3/2}(k\,\eta)}{\left(k\,\eta\right)^{3/2}}\,h^{(\lambda)} ({\bf k},0)\,.
\end{equation}
Using the integral (\ref{integwhite}) along with the above solution yields
\begin{eqnarray}
C_{\ell\ell'mm'} &=& \sum_{\lambda,M,M'}
(-i)^{\ell-\ell'}\,\frac{9\,\pi^3}{16}
\,\sqrt{\frac{(2\,\ell+1)(2\,\ell'+1)(\ell+2)!(\ell'+2)!}{(\ell-2)!(\ell'-2)!}}
\nonumber\\
&&\,\times\left[\left(\delta_{M\,,2}\delta_{M'\,,2}+\delta_{M\,,-2}\delta_{M'\,,-2}\right) \left(\delta^+_\lambda +\delta^\times_\lambda\right) +\left(\delta_{M\,,-2}\delta_{M'\,,2}+\delta_{M\,,2}\delta_{M'\,,-2}\right) \left(\delta^+_\lambda - \delta^\times_\lambda\right) \right]
\nonumber\\
&&
\times\int_0 ^\infty \frac{dk}{k}\,\,
I_\ell^2(k\,\eta_0)\int^{+1}_{-1} \frac{d\xi}{2}\,\int_0^{2\,\pi}\,\frac{d\phi_k}{2\,\pi}\,D^\ell_{m M}({\bf \hat{k}})\,D^{\ell'\,\star}_{m' M'}({\bf \hat{k}})
\,P_{H_\lambda} ({\bf k})\,,
\end{eqnarray}
where $\xi \equiv \cos\theta_k$, the power spectrum is defined through
\begin{equation}
P_{H_\lambda}({\bf k}) = \frac{k^3\,\vert h^{(\lambda)} ({\bf k},0) \vert^2}{2\,\pi^2}\,,
\end{equation}
which coincides with the definitions given in the main text, cf. eqs. (\ref{powtimes}) and (\ref{powepv}) once the universe has isotropized. Finally,    the time integral is
\begin{equation}
I_\ell (k\,\eta_0) \equiv \int_{k\,\eta_{\rm LSS}}^{k\,\eta_0} dy  \,\frac{J_{5/2}(y)}{y^{3/2}}\,\frac{J_{\ell+1/2}(k\,\eta_0-y)}{\left(k\,\eta_0-y\right)^{5/2}}\,.
\label{iell}
\end{equation}
To be able to compare with observations, we average the diagonal ($\ell=\ell'$, $m=m'$) part of the correlator over all $m$ and obtain
\begin{eqnarray}
 C_\ell &=& \frac{1}{2\,\ell+1}\,\sum_m\,C_{\ell\ell m m} 
\nonumber\\
&=& \frac{9\,\pi^3}{8}\,\frac{(\ell+2)!}{(\ell-2)!}\, \int_0^\infty \frac{d(k\,\eta_0)}{(k\,\eta_0)}\,I^2_\ell(k\,\eta_0)\int_{-1}^{+1} \frac{d\xi}{2}\,\int_0^{2\pi}\frac{d\phi_k}{2\,\pi}\left[P_{H_+}({\bf k})+P_{H_\times}({\bf k})\right]\,,
\label{c-ell-app}
\end{eqnarray}
The final result (\ref{c-ell-app}) is valid for any model of pre-inflation which results in an anisotropic primordial tensor power spectrum.

For standard (isotropic) inflation, the power spectrum is the same for both polarizations and we recover the standard result \cite{Starobinsky:1985ww}.


\end{document}